\documentclass[12pt,preprint]{aastex}

\def\msun{{\rm M$_{\odot}$}}
\def\rsun{{\rm R$_{\odot}$}}

\begin{document}

\title{The Effect of Pre-Main Sequence Stars on Star Cluster Dynamics}

\author{Robert Wiersma \altaffilmark{1}, Alison Sills \altaffilmark{2} and
Simon Portegies Zwart \altaffilmark{3}}

\altaffiltext{1}{Department of Physics and Astronomy, McMaster University,
Hamilton, Ontario, Canada, L8S 4M1; current address: Leiden Observatory, P.P. Box 9513, 2300 RA Leiden, Netherlands; {\tt wiersma@strw.leidenuniv.nl}}
\altaffiltext{2}{Department of Physics and Astronomy, McMaster University,
Hamilton, Ontario, Canada, L8S 4M1; {\tt asills@mcmaster.ca}}
\altaffiltext{3}{Astronomical Institute ``Anton Pannekoek'' and Section
Computational Science, University of Amsterdam, Kruislaan 403, 1098 SJ
Amsterdam, Netherlands; {\tt spz@science.uva.nl}}

\begin{abstract}

We investigate the effects of the addition of pre-main sequence
evolution to star cluster simulations. We allowed stars to follow
pre-main sequence tracks that begin at the deuterium burning birthline
and end at the zero age main sequence. We compared our simulations to
ones in which the stars began their lives at the zero age main
sequence, and also investigated the effects of particular choices for
initial binary orbital parameters. We find that the inclusion of the
pre-main sequence phase results in a slightly higher core
concentration, lower binary fraction, and fewer hard binary
systems. In general, the global properties of star clusters remain
almost unchanged, but the properties of the binary star population in
the cluster can be dramatically modified by the correct treatment of
the pre-main sequence stage.
\end{abstract}


\section{Introduction}

Over the past two decades, dynamical simulations of star clusters have
become much more realistic. This realism takes the form of an
increasingly complex treatment of individual stars in the cluster.
For years, dynamical models only considered stars as single equal-mass
points. The introduction of a mass function into dynamical models
quickly necessitated some treatment of stellar evolution. High mass
stars have much shorter lifetimes than low mass stars, mass loss from
high mass stars can remove a significant fraction of the mass from the
cluster, and very high mass stars can completely dominate the
dynamical evolution of the cores of clusters \citep{EHI87, MOD1,
PBHMM04}. Binary stars also have a substantial impact on the cluster
dynamics, by acting as energy sinks or sources. It has been known for
a long time that a single massive binary can dominate the evolution
of star clusters \citep{A71}. A hard binary system can produce cluster
energy through an encounter with another system that leaves the binary
system with a tighter orbit. Encounters between soft binary systems
and other objects can release the binary's binding energy to cool the
cluster, whereas interactions with hard binaries effectively heat the
cluster. Dynamically produced binaries were recognized as a key
population for halting core collapse \citep{EHI87}. Open clusters also
have primordial binaries \citep[][e.g.]{ML86,Keta05}, and those binary
systems can affect the cluster evolution from its birth.

Stellar dynamicists realized that the point mass approximation for
stars was neglecting a number of dynamically significant processes in
clusters. Allowing stars to have radii that change as they evolve was
an important next addition to stellar dynamics simulations
\citep{PMHM01,HTAP01,dlFM02}. Finite stellar radii are most
important for two aspects of these simulations. First, binary stars
can undergo mass transfer as one member of the system fills its Roche
lobe, either through evolution of the star or dynamical modification
of the orbital parameters of the system. Changing binary systems will
change how the binaries affect the evolution of the cluster. In the
extreme case, the two components of the binary system can merge.
Secondly, stars with finite radii can collide with other stars, either
through direct hyperbolic collisions (in dense clusters) or in highly
eccentric binary systems created after close encounters. One of the
earliest papers to show the dynamical importance of binary-single
encounters was written 30 years ago \citep{H75}.  Stellar collisions
can produce blue stragglers \citep{Seta97} and other non-standard
stellar populations \citep{PMMH99}. These populations can in turn
modify the dynamical evolution of the cluster.

Young, open clusters are of particular interest when simulating star
clusters. Because they tend to have a relatively small number of
stars, open clusters provide observations that do not require
prohibitive computational expense to reproduce. They also are well
studied and the evolution of their stellar populations (population I)
are well determined. Some open clusters are young enough to have
pre-main sequence stars that are observable; indeed numerous authors
have reported pre-main sequence stars in the Pleiades for instance
\citep{GRM94,Ple03}. For dynamicists, open clusters showcase a variety
of phenomenon that play an important role in the evolution of the
cluster. The environment that typifies the core of such a cluster is
dense enough to provide for a rather high (and well known) binary
fraction, and collisions, mergers, and other encounters are expected.
All of these factors make open clusters an excellent place to start
when integrating stellar evolution with dynamics.

A number of authors have endeavored to push the limits of realism in
dynamical simulations. \citet{HTAP01} start their simulations with
10000 - 15000 stars distributed via Plummer and King models with
varying fractions of binaries. They assign masses to these using
stars using \citet{KTG91,KTG93} initial mass functions, and
evolve them from the zero age main sequence using recipes detailed in
\citet{Heta00}. Their model includes the effects of the galactic tidal
field. Because their study focused on blue stragglers, they included
collisions and mergers as well as close encounters. \citet{PMHM01}
(hereafter PZ01) start with 3072 stars all distributed using a King
model, and using a \citet{S86} mass function. Their treatment of
binary and stellar evolution is somewhat different from
\citet{HTAP01}, and will be discussed in more detail in later sections
of this paper.

All previous work assumed that all stars began their lives on the zero
age main sequence. However, low mass stars make up the bulk of stars
in a cluster for any reasonable initial mass function. Low mass stars
also have a pre-main sequence lifetime that is a substantial fraction
of the cluster lifetime, significantly influencing a cluster
population. These young stars have radii which can be up to 10 times
larger than their main sequence radii. Therefore, some binaries could
have undergone an episode of mass transfer that was hitherto not taken
into account. Also, larger stars are more likely to have experienced a
collision; those collision products will have been missed in previous
simulations. Including pre-main sequence evolution is regarded as a
potentially important step in increasing the realism of stellar
dynamics simulations
\citep{MOD2}.

In this paper, we explore the effect of including the pre-main
sequence phase of stellar evolution in a stellar dynamics calculation.
We look at the differences between simulations with and without the
pre-main sequence in terms of global cluster properties (density
profile, dissociation time, etc.) and in terms of the observable
effect on the colour-magnitude diagram of the cluster. We look at the
change in number and nature of unusual stellar populations and of the
evolved stellar populations. In section 2, we outline the dynamical
method used and how we incorporated pre-main sequence evolution. In
section 3 we present our results, and discuss their implication in
section 4.

\section{Method}

For all our simulations, we use the STARLAB software environment,
featuring the {\tt kira} integrator \citep{M96,PMHM01} and {\tt SeBa}
stellar and binary evolution package \citep{PZV96,PZY98}. The
simulations were run using the GRAPE-6 hardware \citep{MFKN03}.

\subsection{Initial Conditions}

In the interests of simulating observable results, we choose
parameters corresponding to population I, young clusters. These
clusters typically have approximately 1000 stars, and are no older
than a few billion years.

We ran three sets of three simulations with different realizations of
the input parameters. Our initial conditions mirror those of PZ01,
except that in some cases, our stars begin on the pre-main sequence
rather than the zero age main sequence. In order to effectively
determine the degree that adding pre-main sequence evolution affects a
cluster simulation, we use the same input snapshots for the first
three W6 models of PZ01, with various modifications outlined below.

For the input snapshots, PZ01 begin the simulation with 2048 nodes
(single or binary stars) set up with a \citet{K66} model with a
dimensionless depth ($W_0$) of 6, taking into account the velocity
anisotropy and non-spherical shape that cluster would experience in
the Galactic tidal field similar to that of \citet{HR95}. The initial
virial cluster radius was taken to be 2.5 pc. They then add a binary
companion to every second node, for a total of 3072 stars. This is
comparable to young clusters such as NGC 2516 and the Pleiades. Masses
were applied to the nodes using a initial mass function prescribed by
\citet{S86} with masses ranging from 0.1 \msun\ to 100 \msun, and a
mean mass $\langle m\rangle \simeq$ 0.6\msun. This yields an initial
total mass of $M_0 \sim 1600$\msun, which is similar to estimates for
the Hyades of \citet{W93}. The masses for the secondary stars were
selected randomly between 0.1 \msun\ and the mass of the primary. The
orbital eccentricities were selected from a thermal distribution
between 0 and 1. The orbital separation $a$ was selected with a
uniform probability in log $a$, with a minimum separation of Roche
lobe contact or 1 \rsun, whichever is smaller. The maximum separation
of the binaries was taken to be $10^6$ \rsun (about 0.02 pc). Note
that the condition of Roche lobe contact implicitly includes some
dependence the initial radius of the star. Finally, each star was
assigned a radius, luminosity, and temperature based on their current
evolutionary state. Main sequence initial prescriptions for radius,
temperature, and luminosity are given by \citet{EFT89}. We added
pre-main sequence attributes (where applicable) using the tracks
obtained from \citet{SDF00}.

We labeled our simulations as `pz-ms', `pz-pms' or `rw-pms', with
differences as follows. For our pz-ms runs, the snapshots were
identical to those used in PZ01 -- the positions, velocities, masses
and other stellar and binary parameters were not changed. These runs
were used for comparison purposes, since all stars in this set of
simulations began their lives on the zero age main sequence. For our
pz-pms runs, the positions, velocities, masses, and binary parameters
were the same as those used in PZ01 but all low mass stars ($M \leq
7$\msun) were started on the pre-main sequence birthline rather than
on the zero age main sequence. Essentially the only initial condition
that differed between the pz-ms and pz-pms runs was the stellar radius
for low mass stars. The condition for chosing binary orbit separations
was a limit based on Roche lobe contact, but pre-main sequence stars
have much larger radii than zero age main sequence stars. Therefore,
many of the binary systems in the pz-pms simulations were initially in
contact. To address this issue, we also ran a third set of simulations
(rw-pms) in which the positions, velocities and masses of stars were
identical to PZ01, but the binary semi-major axes were re-assigned,
with a limit based on Roche lobe contact as determined by the pre-main
sequence radii. For each set of initial conditions, we ran three
different realizations, corresponding to the first three realizations
reported in PZ01. Where relevant, we label individual realizations as
`pz-pms1', etc.

Adding the pre-main sequence phase of evolution to the stars in our
cluster highlights some confusion in dating clusters. What exactly is
meant by an age of zero? If the pre-main sequence phase is included,
then low mass stars will reach the zero age main sequence at different
times, all of which are later than the start of the simulation, which
we define as $T=0$. In PZ01, it was clear that $T=0$ corresponded to
the moment when all stars are on the zero age main sequence. For our
pre-main sequence simulations, we take $T=0$ to be the moment when all
stars are on the deuterium-burning birthline as defined by
\citet{PS99} (see section \ref{PMSev}). The difference between these
two implementations is dependent on the mass of star in question. When
comparing ages from dynamical simulations to those determined from
observations, it is important to understand how the observed ages were
derived.

\subsection{Evolution}

The entire system is evolved, including the effects of dynamics,
individual stellar evolution, and binary evolution. For a complete
description of all the computational techniques, the reader is
directed to the descriptions of STARLAB, {\tt kira}, and {\tt SeBa}
\citep{M96,PMMH99,PZV96,PZY98}. In this section, we will concentrate
on describing the numerical methods and parameters that were modified
when we introduced the pre-main sequence phase of evolution.

Single star evolution from the zero age main sequence is based on
the time-dependent mass-radius and mass-luminosity relations given by
\citet{EFT89}. These relations are valid for the evolution of
solar-metallicity stars on the main sequence, subgiant branch or
Hertzsprung gap, giant branch, horizontal branch and asymptotic giant
branch. Stars are immediately turned into inert remnants (white dwarf,
neutron star or black hole) at the end of the asymptotic giant
branch. The effective temperature and bolometric correction can be
calculated for each star from the mass, radius and
luminosity. Quantities such as core mass and mass loss due to a
stellar wind are also parameterized, as outlined in \citet{PZV96}.

The evolution of the stars and binaries are computed with the {\tt
SeBa} binary population synthesis package (see Portegies Zwart \&
Verbunt, 1996, Sect.\, 2.1\nocite{PZV96} with changes made in model B
of Portegies Zwart \& Yungelson, 1998).\nocite{PZY98} In this
synthesis model the two stars in a binary are evolved synchronously
updating each of the stars at regular intervals while keeping track of
the changes to the orbital period and eccentricity. The simulation
model includes various stability criteria for mass transfer, tidal
circularization, the emission of gravitational waves, supernova
explosions, etc.

\subsection{Treatment of Pre-main Sequence Stars \label{PMSev}}

\begin{figure*}
\plotone{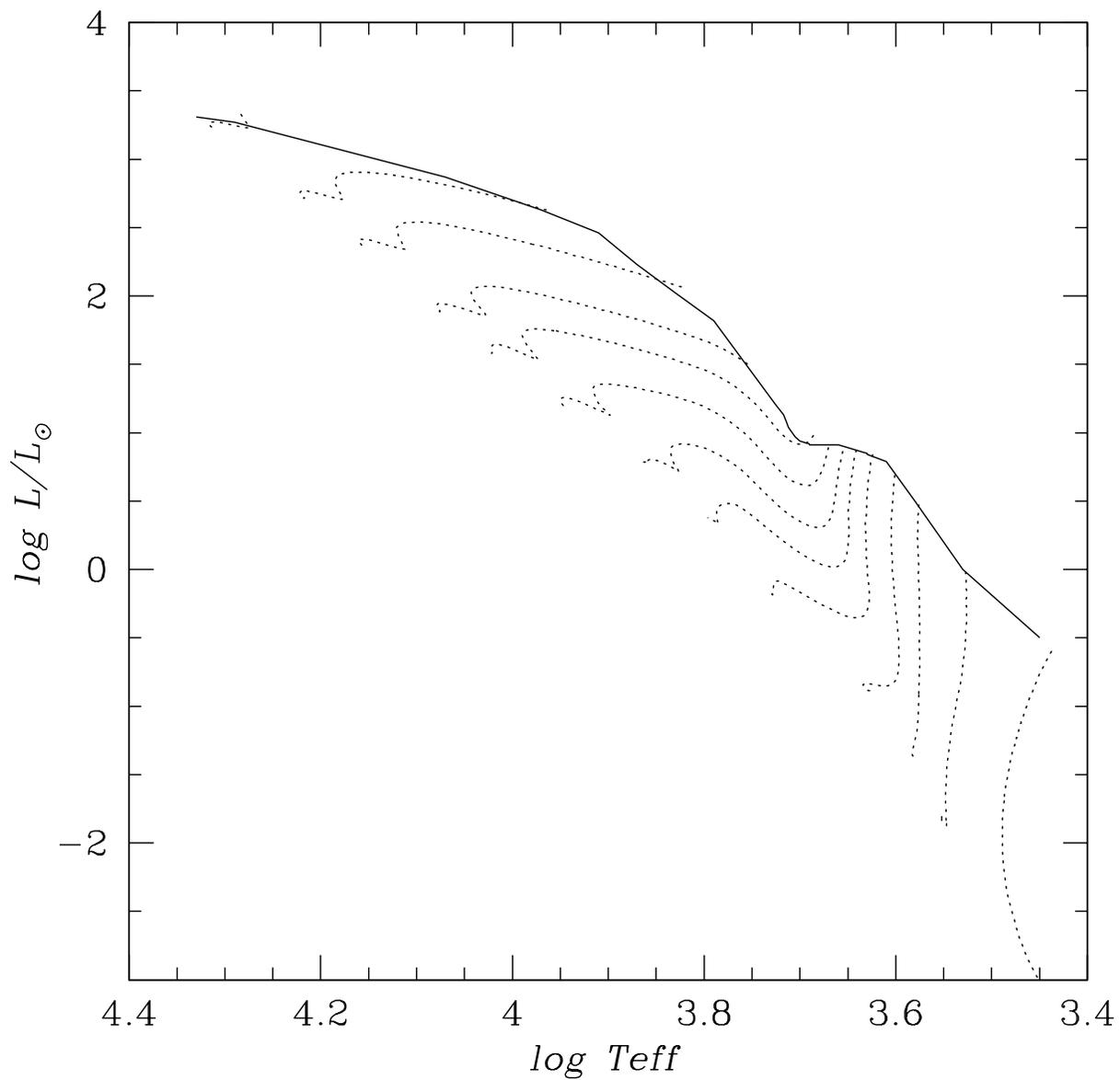}
\caption{Pre-main sequence tracks of \citet{SDF00}, showing the region
between the birthline of \citet{PS99} (solid line) and the zero age
main sequence. The masses shown are 0.1, 0.3, 0.5, 0.7, 1.0, 1.3, 1.6,
2.0, 2.5, 3.0, 4.0, 5.0 and 7.0 \msun\label{PMStracks}}
\end{figure*}

For our pre-main sequence recipes, we looked to the evolutionary
tracks of \citet{SDF00}, a selection of which are shown in figure
\ref{PMStracks}. They give pre-main sequence tracks that begin with a
fully convective protostar, and end just after hydrogen burning
begins. Their tracks have resolution in mass of 0.1 \msun\, from 0.3 to
2 \msun, higher mass resolution between 0.1 and 0.3 \msun, and lower
resolution between 2 and 7 \msun. We used the tracks for a metallicity
of $Z = 0.02$ (i.e. solar metallicity, appropriate for young open
clusters) and used the portions of their tracks which begin at the
pre-main sequence birthline described by \citet{PS99} and end at the
zero age main sequence. We did not implement the pre-main sequence
phase for stars with masses greater than 7 \msun\ because these stars
spend little or no time on the pre-main sequence. We combined all the
tracks to create a lookup table in which we interpolated linearly
through mass and age to find radius and temperature.

\clearpage

\clearpage

Although other pre-main sequence tracks exist \citep{DM97, PS99}, we
chose to use the \cite{SDF00} tracks. They have sufficient resolution
and span the entire range of masses to which pre-main sequence
evolution applies. Since theoretical tracks from different authors
differ amongst themselves it was important to be able to have enough
data to infer the properties of the star for a given mass and age. The
range of masses is quite appropriate since our initial mass function
has a lower limit of 0.1 \msun, and stars with masses higher than
about 7 \msun\, do not have a significant pre-main sequence phase.
Another reason we chose these tracks was that they had data for
stellar radius and temperature that were easily
manipulated. \citet{SDF00} compare their tracks with others and find
good correspondence with most other tracks, although there is still
some debate concerning the form of very low ($M \leq 0.4$ \msun) mass
pre-main sequence evolution.

Beyond temperature, radius and luminosity, a number of other
parameters need to be specified for pre-main sequence stars. These
parameters come in two categories: those dealing with single-star
evolution, and those dealing with binary evolution. Unfortunately,
many of these parameters have not been determined for pre-main
sequence stars. To make reasonable choices, we had to look to stars
with similar composition (main sequence) or similar structure (the
large convective envelope of a giant) to give us these other
characteristics. For instance, the way that we chose to calculate the
mass transfer and age rejuvenation from accreting binary star systems
in STARLAB is similar to the main sequence recipes, but the parameters
controlling mass transfer relations are based on the values for giant
stars. Similarly, we took the value for the radius of gyration to be
that of giant stars. Since pre-main sequence stars do not exhibit mass
loss (especially stars with masses less than 7 \msun), we neglected
stellar winds on the pre-main sequence.

One of the major benefits of using the STARLAB environment is that
there are prescriptions for mergers or collisions of varying types of
stars. Collisions and mergers can be distinguished in that a merger
is a result of unperturbed binary evolution, and is preceded by a
period of mass transfer. Currently, the code does not treat
collisions that do not result in a merged object, so at the moment when
the two stars are replaced by a single star the result will depend
only on the types of stars and their mass.

Encounters involving main sequence and post-main sequence stars are
described in PZ01. The merger of two main sequence stars is treated as
conservative mass accretion from the less massive secondary to the
more massive primary. This results in a rejuvenation of the star --
which usually is observed as a blue straggler if the masses of the
stars are large enough, or as a reasonably normal main sequence stars
if the total mass of the new object is less than the current turnoff
mass. The merger of other types of stars result in evolved stars or
unusual objects, as warranted by the structure of the two stars
involved in the collision.

The treatment of encounters involving pre-main sequence stars was
similar to the treatment of main sequence stars -- when a pre-main
sequence star collided with a more evolved star, the result was taken
to be similar to the result of a main sequence star colliding with the
same kind of star. For pre-main sequence/pre-main sequence collisions,
the merged object was returned to the pre-main sequence birthline as
its evolutionary state will be completely disrupted. These choices are
in agreement with the results of the hydrodynamic simulations of
\citet{LS05} of collisions involving pre-main sequence stars.

\section{Results}

\subsection{Global Cluster Properties}

Two of the most general functions that represent the time evolution of
star cluster are its total mass and total number of particles versus
time. The processes which can decrease the total number of cluster
members are a merger between two stars, type Ia supernovae (which do
not leave a remnant) and the escape of a star from the system. Stars
may escape as a result of a mixture of influences: dynamical encounter
with a binary, supernova kick, or removal by the galactic tidal field.

\begin{figure*}
 \plottwo{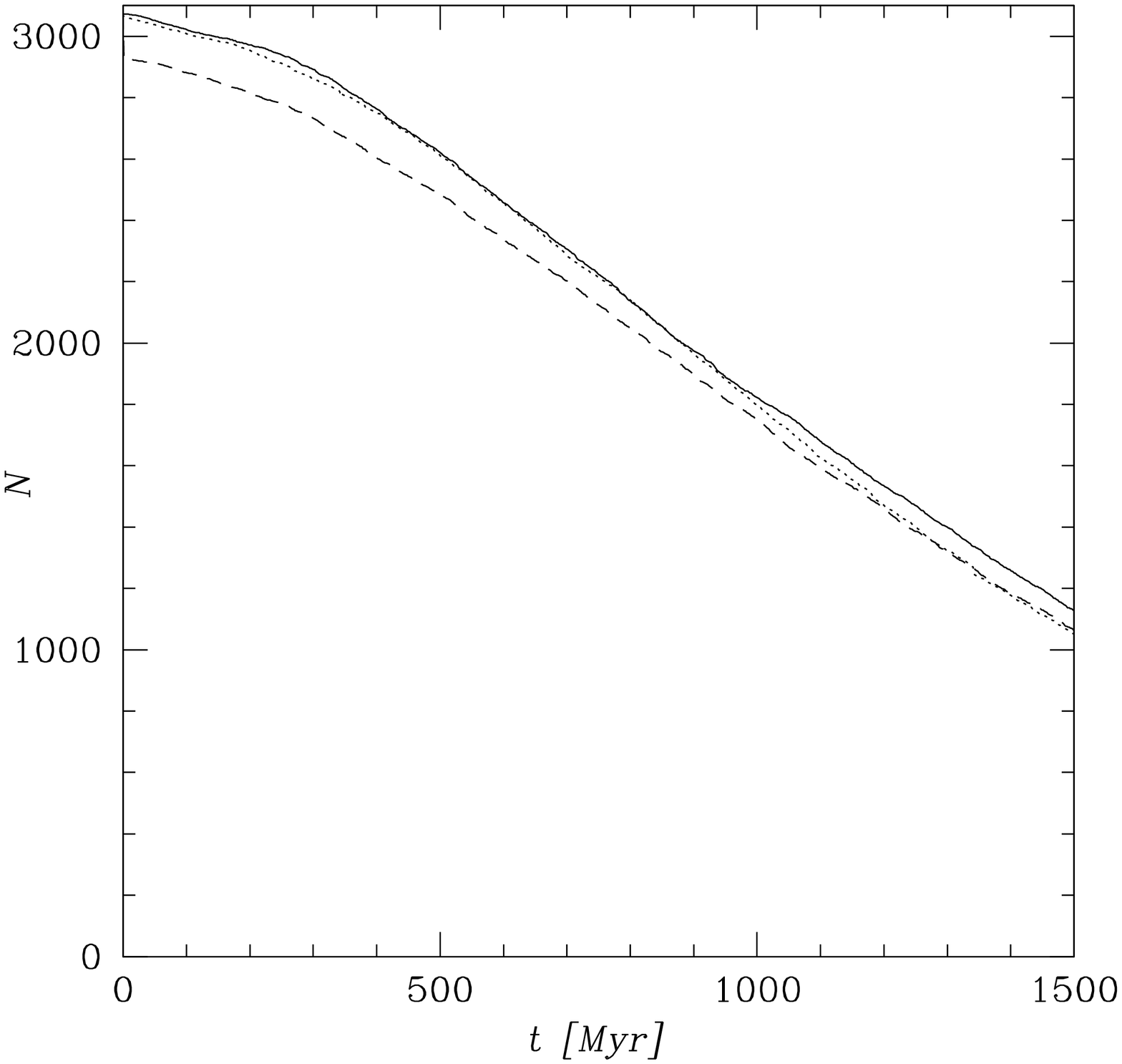}{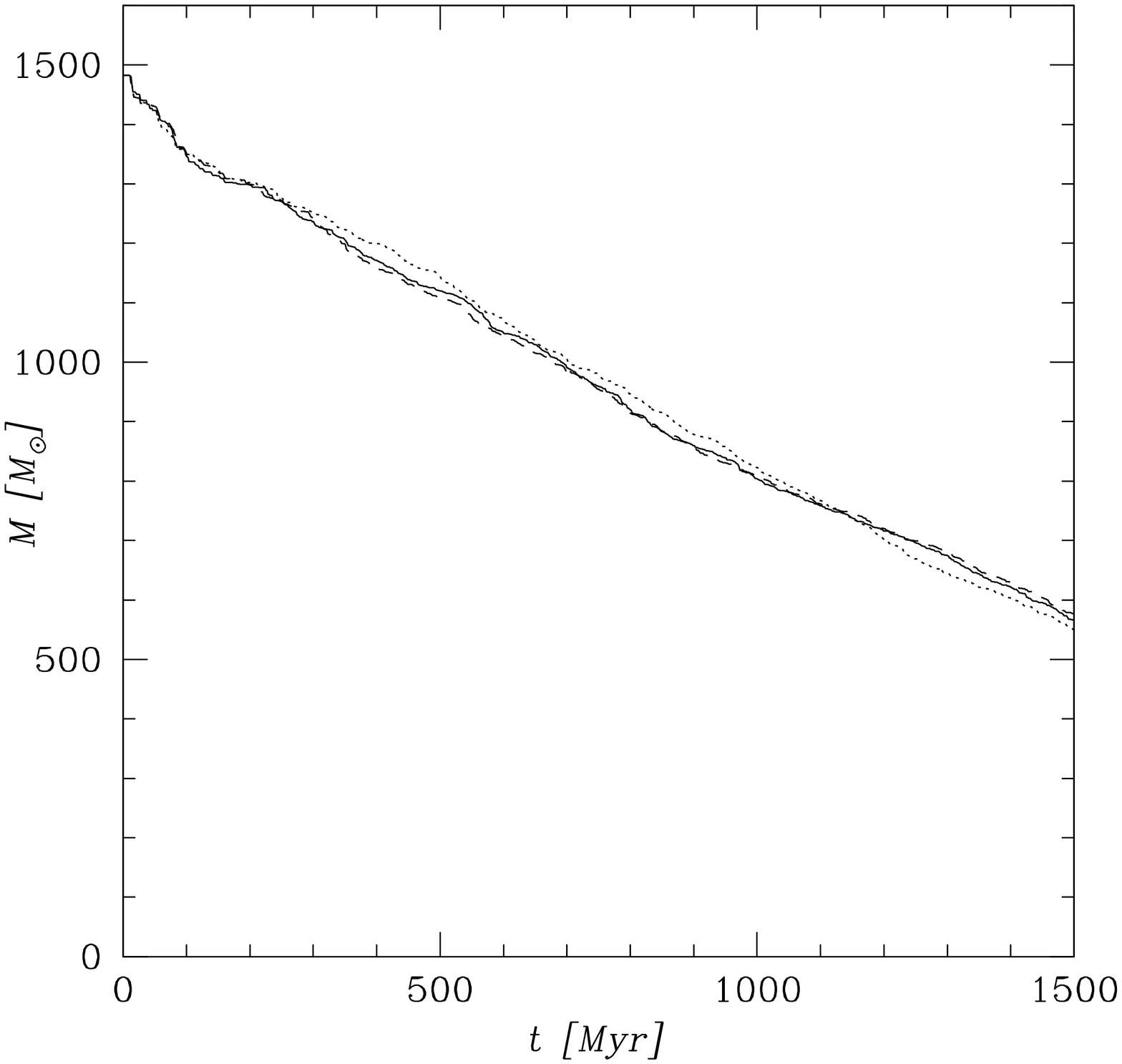}
\caption{Total number of stars in the cluster averaged all runs (left),
and total mass of run 1 (right). Results shown are from the pz-ms (dotted),
pz-pms (dashed), and rw-pms (solid) sets of simulations.\label{mnvt}}
\end{figure*}

Figure \ref{mnvt} shows the total number of stars (where a binary
system counts as 2 stars) versus time for each set of runs. The
largest effect is the large initial drop in the number of stars in the
pz-pms runs (dashed line). This has mainly to do with early mergers
that occur within these clusters. In these runs, all of the binaries
that had a small orbital separation in the main sequence configuration
are now in contact. As a result, a large number of binaries merge
immediately after initialization of the simulation. Eventually, the
evidence of this event is erased, and the number of remaining cluster
members approaches that of the other series.  The rw-pms runs
experience a decreased number of mergers throughout the life of the
cluster since the radii of the stars become smaller throughout the
pre-main sequence lifetime, and may only reach a point of Roche lobe
overflow when one of the stars becomes a giant. As a result, the rate
of decrease of the number of stars for the rw-pms runs ($-1.5$
stars/Myr) is lower than the rate the pz-ms runs ($-1.6$ stars/Myr),
but with a very similar slope to the pz-pms runs ($-1.44$
stars/Myr). The slopes were measured between 500 and 1500 Myr.  This
dependence of the dynamical results on the initial binary merger rate
highlights the most important conclusion of this paper. The inclusion
of pre-main sequence stars in dynamical simulations affects only the
binary properties of the system, and a consistent treatment of the
pre-main sequence binary population is necessary.

The evolution of the total mass of the cluster is driven by two
processes. One is mass loss from stars via stellar winds, and the
other is the escape of stars from the cluster. The differences caused
by including the pre-main sequence phase are quite small, as shown in
figure \ref{mnvt}. The maximum difference between the total mass of
the pz-ms run and both pre-main sequence runs is at most 6\%.

\clearpage

\begin{table}[t]
\caption{Number Loss and Mass Loss at 1.5 Gyr for all runs}
\label{massloss}
\centering
\begin{tabular}{lrrr}
\hline\hline
 & pz-ms & pz-pms & rw-pms \\ \hline
\% of stars that have escaped & 65 $\pm$ 4 & 60 $\pm$ 5 & 60 $\pm$ 10 \\
\% objects that have merged & 3.1 $\pm$ 0.5 & 15 $\pm$ 2 & 0.4 $\pm$ 0.2 \\
\% total mass lost via escapers & 48 $\pm$ 4 & 45 $\pm$ 5 & 45 $\pm$ 5 \\
\% total mass lost via stellar winds & 19 $\pm$ 3 & 20 $\pm$ 3 & 19 $\pm$ 3 \\ \hline
\end{tabular}
\end{table}

\clearpage

Table \ref{massloss} summarizes which modes of mass loss and number
loss are experienced in each of the models. The error bars on each
number give an indication of the range between different realizations
of each set of initial conditions. Within the errors, the same number
of stars escape from the cluster in all runs since the escape
processes are driven by stellar dynamics, which is most influenced by
the masses of the stars and binaries, and which is hardly affected by
the pre-main sequence evolution. As noted above, the number of mergers
is much higher in the pz-pms runs because of the combination of the
initial orbital semi-major axes and the larger pre-main sequence
evolutionary radii. These mergers cause the average individual stellar
mass to increase slightly. However, there is no noticeable change in
the amount of mass lost through stellar winds because stars that
exhibit the pre-main sequence phase have masses of less than 7
\msun. Therefore, the high-mass stars which exhibit significant mass
loss ($M \gtrsim$ 25 \msun) will have their masses increased by a very
small amount, which will not dramatically change the amount of mass
loss due to stellar evolution for the cluster. Recall that high-mass
stars do not have pre-main sequence evolutionary tracks, and therefore
are unaffected by the changes made in this paper.

\clearpage

\clearpage

\begin{figure*}
\epsscale{.80}
\plotone{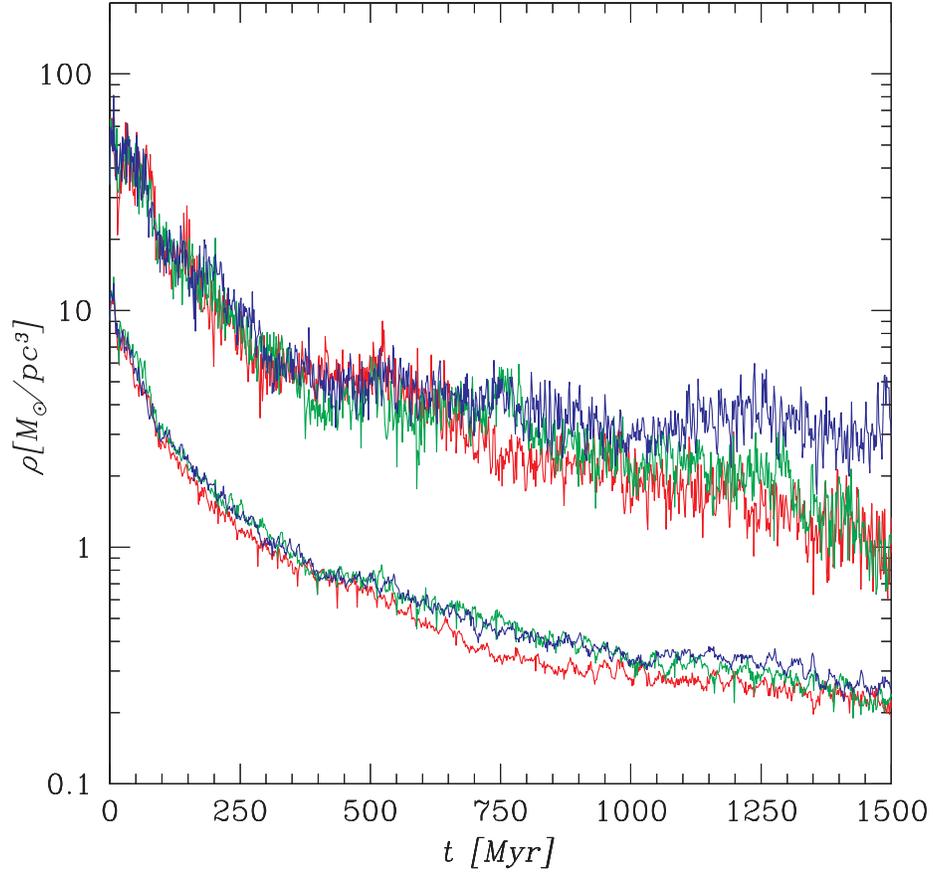}
\caption{Stellar density versus time averaged over all realizations of
models pz-ms (red), pz-pms (green), and rw-pms (blue). The top and
bottom lines represent the density within the 10\% and 50\% Lagrangian
radii respectively.\label{den}} %colour figure
\end{figure*}

Figure \ref{den} shows how density in the core and at the half-mass
radius of the cluster changes with time for the different models. The
overall density of the clusters seem to stay the same, but at around
1.2 Gyr, the cores of the pz-ms and pz-pms clusters continue to
decrease in density, while the density in the core of the rw-pms runs
becomes approximately constant. This may indicate that the core of
these clusters can still absorb energy as massive stars fall toward
the center and low mass stars are ejected, and is related to the
number and hardness of binaries in the core \citep{2004ApJ...608L..25M}.

\clearpage

\clearpage

\begin{figure*}
\epsscale{0.4}
 \plotone{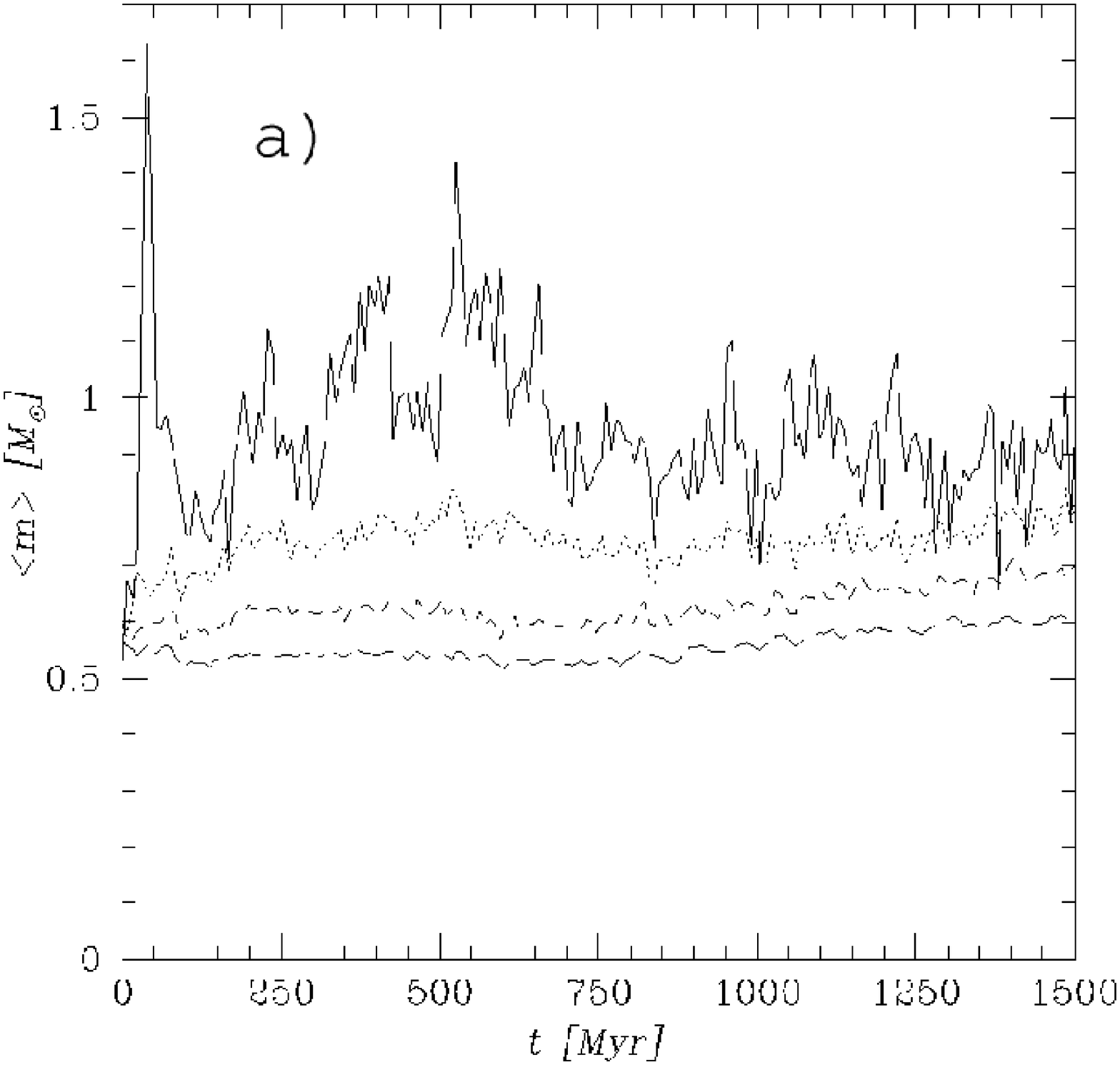}
 \plotone{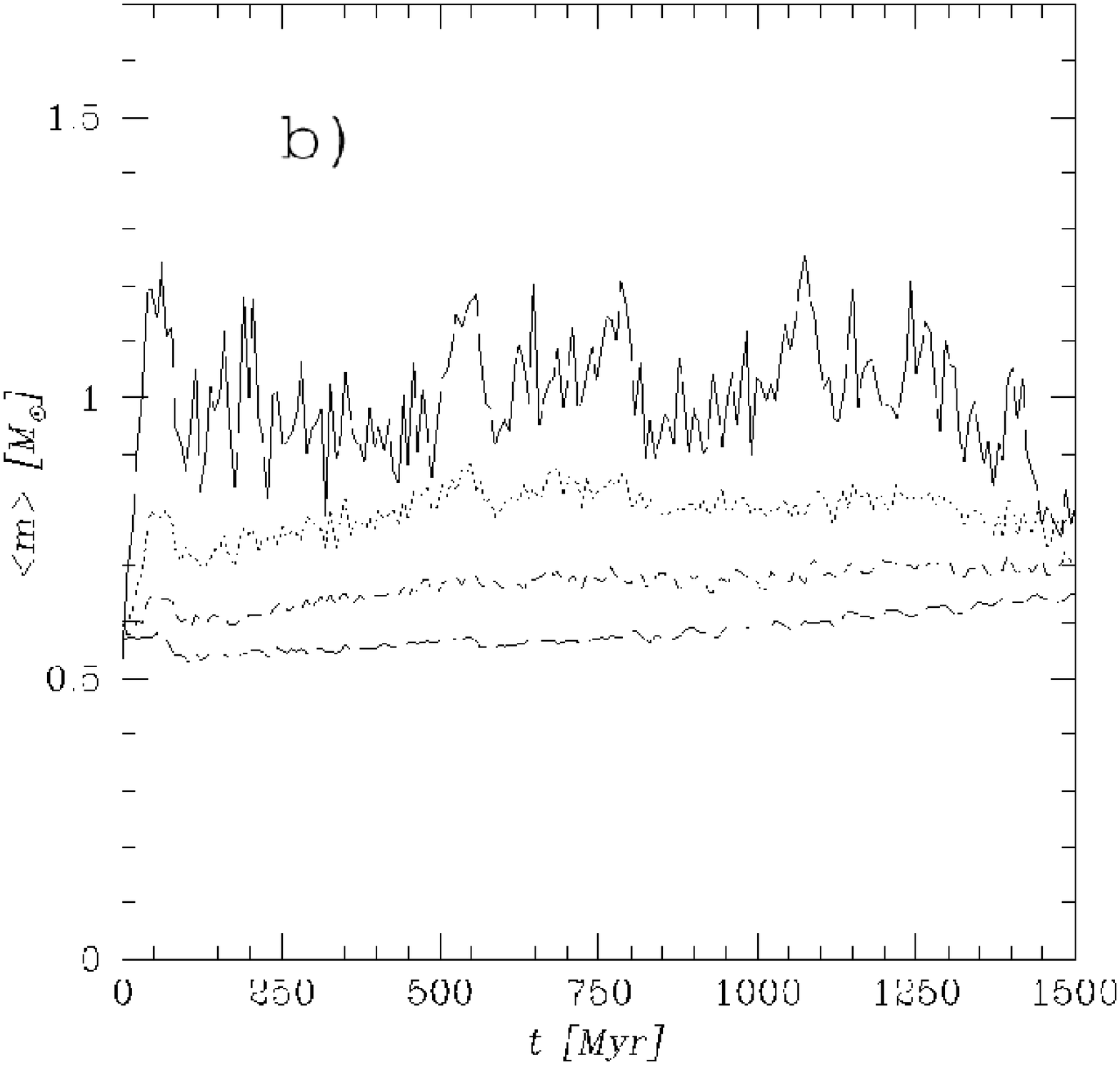}
 \plotone{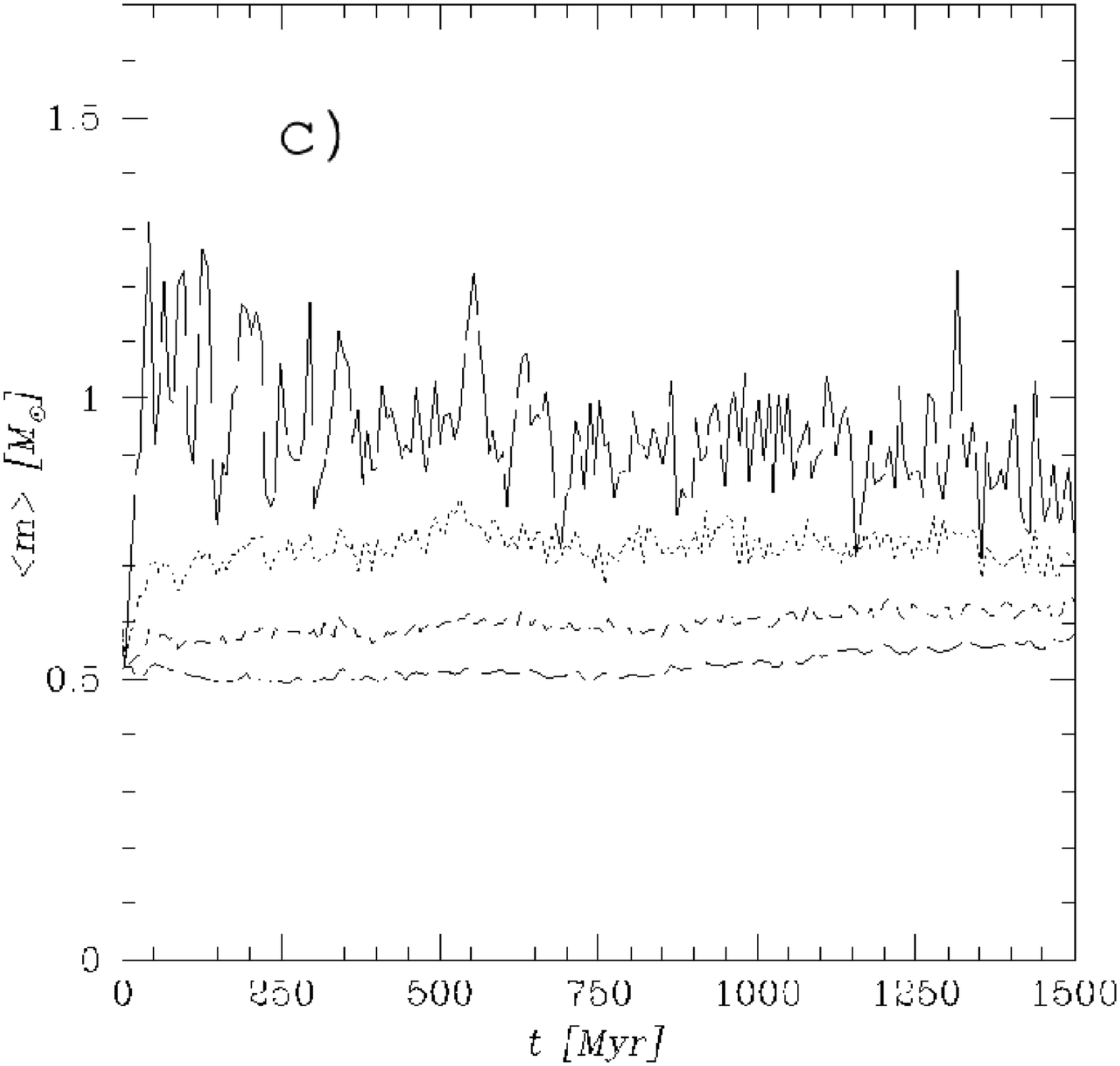}
\epsscale{1}
\caption{Average stellar mass versus time for pz-ms1 (panel a), pz-pms1
(panel b), and rw-pms1 (panel c). Shown are the average mass calculated
inside the 5\% (top line), 25\%, 50\%, and 75\% (bottom line)
Lagrangian radii. The values were smoothed over 7.5 Myr
intervals.\label{avm}}
\end{figure*}

We see evidence for mass segregation in all our runs, as seen in
figure \ref{avm}, a plot of the average stellar mass within various
Lagrangian radii as a function of time. The average mass in the pz-pms
and rw-pms runs increase dramatically at first, and then level off to
a value slightly above the pz-ms runs. The average mass inside the 5\%
Lagrangian radius, averaged between $T=0$ and $T=1500$ Myr, for the
three runs are 1.00 $\pm$ 0.07 \msun\, for the pz-ms run and 1.00 $\pm$
0.04 \msun\, for the pz-pms run but 1.2 $\pm$ 0.1 \msun\, for the rw-pms
run. Again, the initial mergers that the pre-main sequence stars
experience result in masses higher than normal, yielding a higher
average mass especially in the innermost region of the cluster.

\clearpage

\clearpage

\begin{figure*}
\plottwo{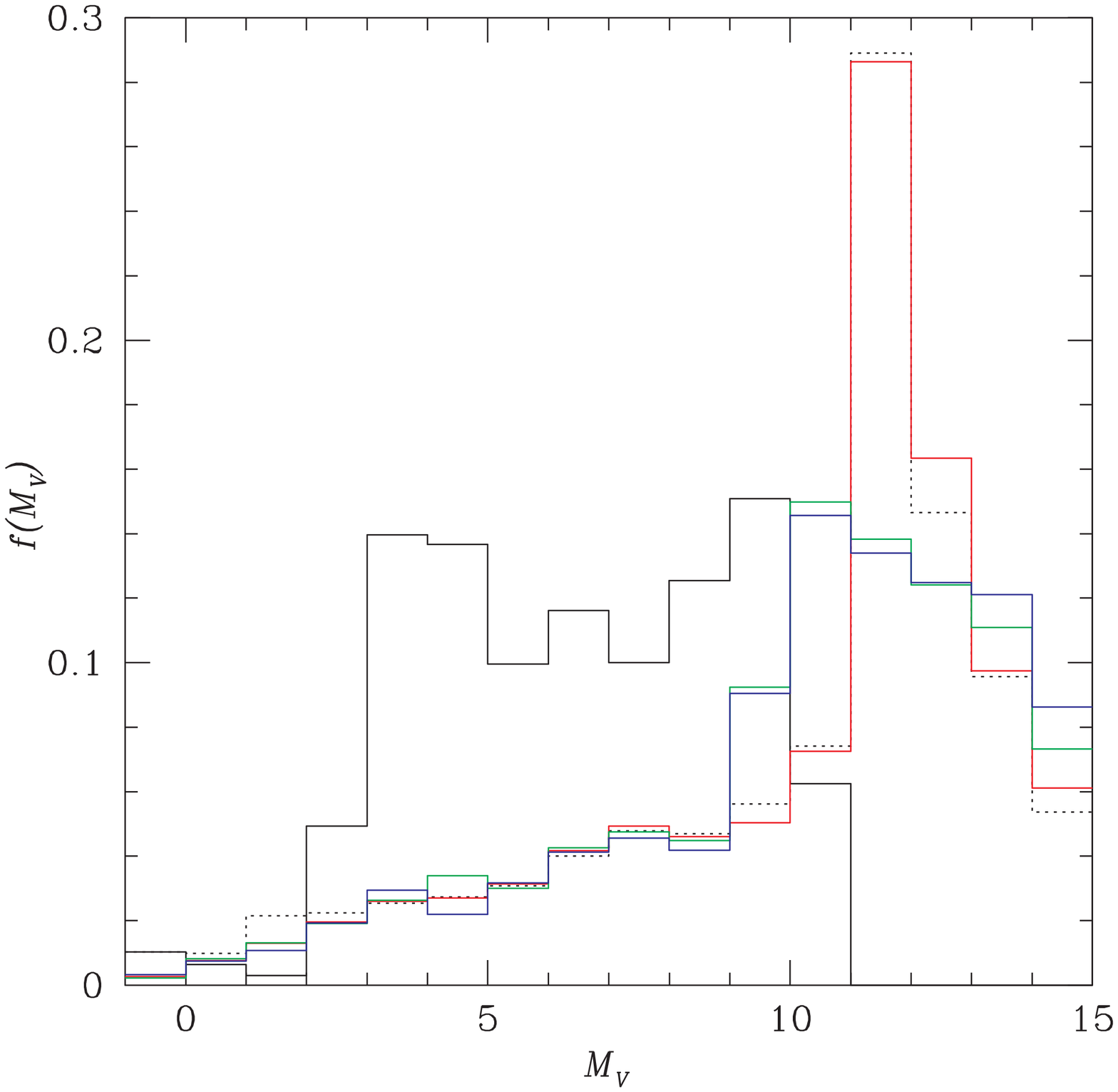}{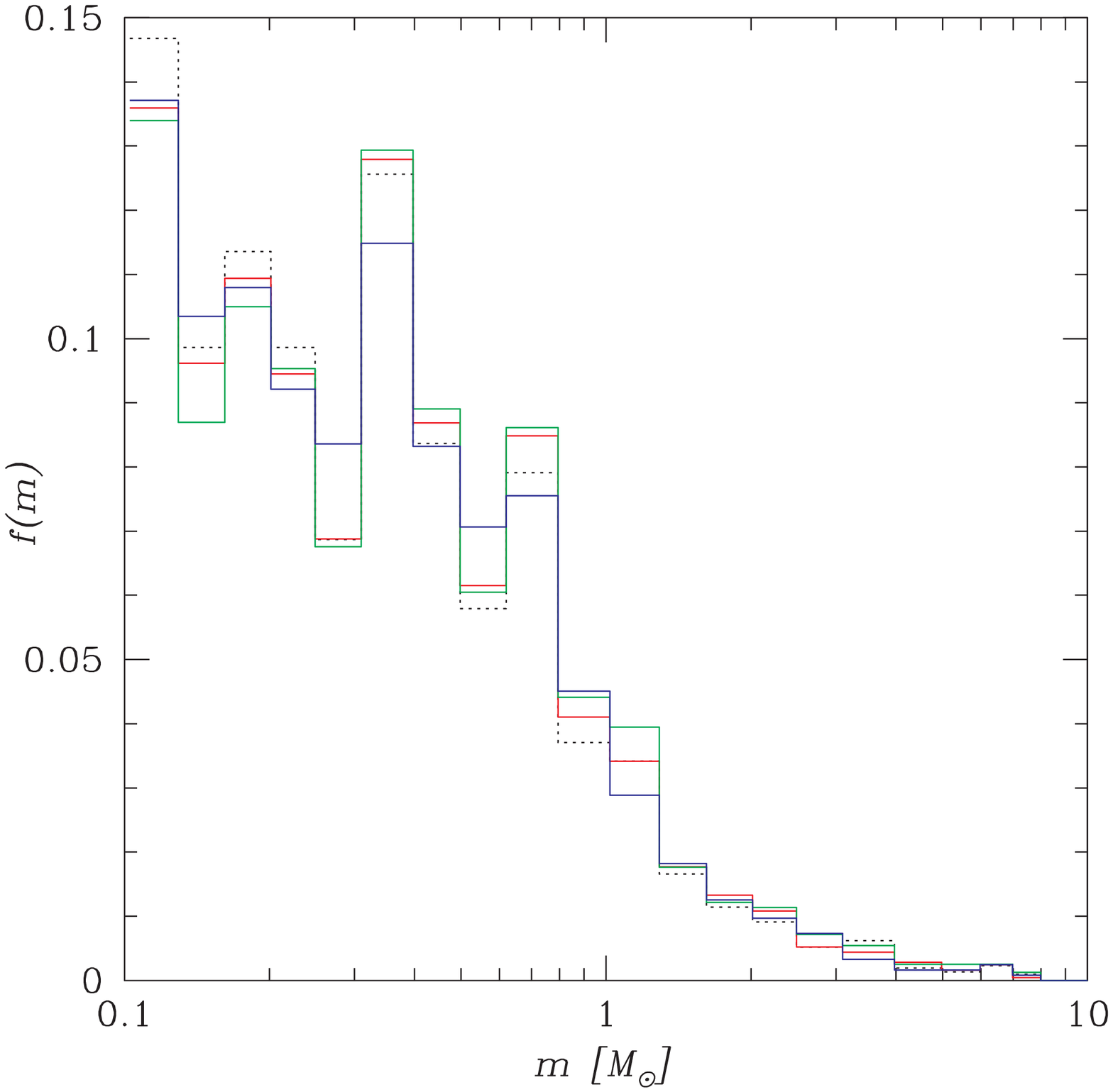}
\caption{Luminosity (left) and mass (right) functions for models pz-ms1,
pz-pms1, and rw-pms1. The dotted line represents the initial
configuration for the pz-ms run (which is identical to the models
starting on the pre-main sequence for the mass function) and the solid
black line represents the initial configuration for the pz-pms run and
the rw-pms run. Also shown are the data at 600 Myr for pz-ms (red),
pz-pms (green), and rw-pms (blue).\label{lfmf}} %colour figure
\end{figure*}

Figure \ref{lfmf} shows the evolution of the mass and luminosity
functions over time in our simulations. The runs in which the stars
begin on the pre-main sequence have an initial luminosity function
which is strongly weighted towards bright stars, as expected. After
600 Myr these models still do not fill the low luminosity bins because
the lowest mass stars still have not evolved onto the main
sequence. In the mass function diagram, the pz-ms1 and rw-pms1 models
evolve similarly, with the pz-pms1 model having fewer low mass stars
(mainly due to a high number of early mergers). A Kolmogorov-Smirnoff
test cannot distinguish between the various mass functions at the 97\%
level. The luminosity functions of the two pre-main sequence runs
(pz-pms and rw-pms) after 600 Myr are drawn from the same distribution
to within 99.8\%, while the pz-ms and rw-pms luminosity functions give
a KS probability of 89\% (i.e. only marginally different). As is
expected, the two initial luminosity functions have only a 13\%
probability of being drawn from the same distribution. Therefore, it
would appear that after only 600 Myr, most of the observational
differences between the two treatments of starting point for the
cluster stars have been erased. This is not a surprise, since the
cluster is a few initial half-mass relaxation times old, so one would
expect that small changes to the initial conditions or to the early
stellar evolution have been erased.

\clearpage

\clearpage

\begin{figure*}
\plotone{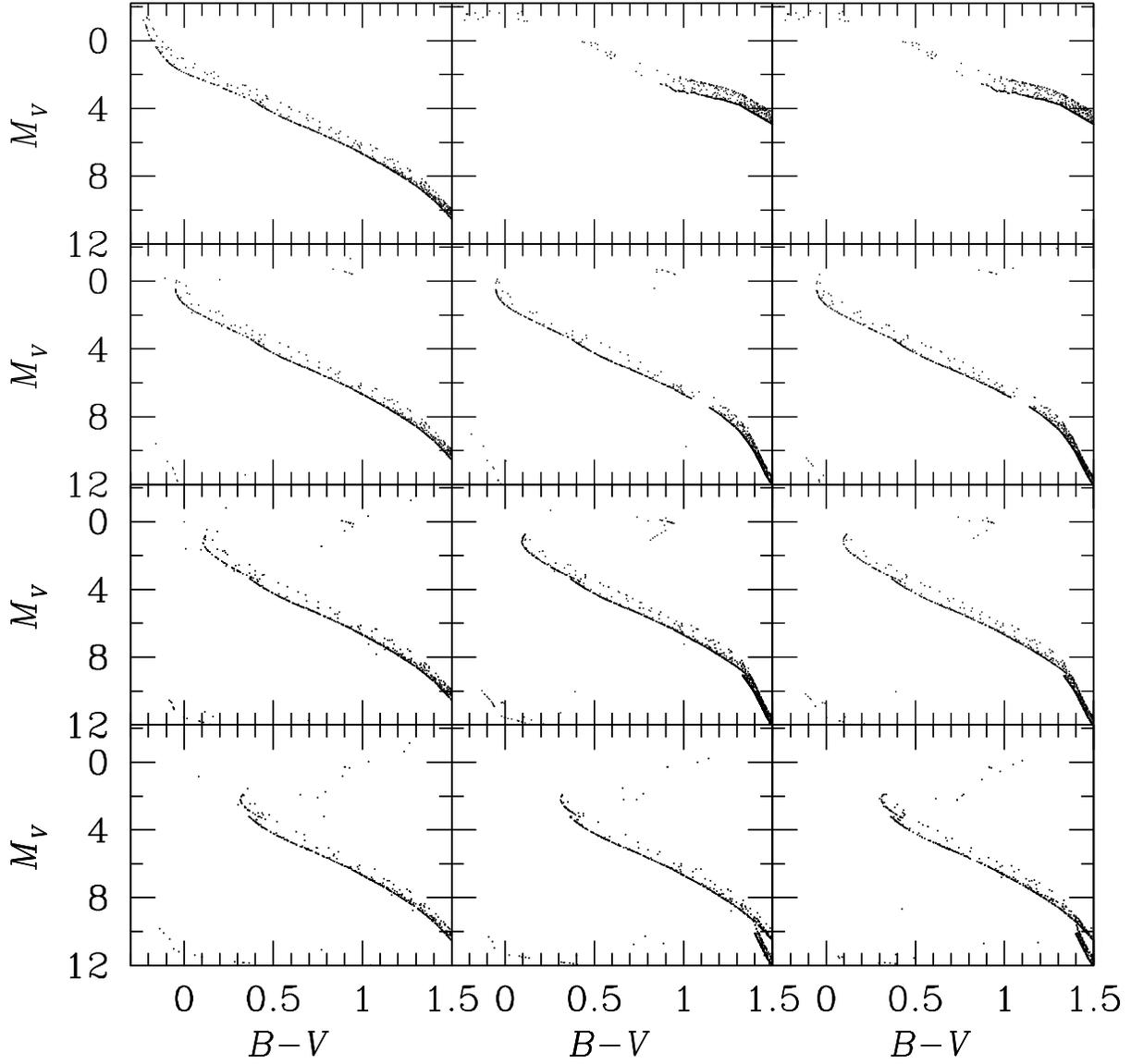}
\caption{Colour-Magnitude Diagrams (CMDs) for models pz-ms1 (left),
pz-pms1 (center), and rw-pms1 (right). Descending chronologically, the
CMDs shown represent the clusters at approximately 0, 300, 600, and
1200 Myr.\label{fig-cmds}}
\end{figure*}

In figure \ref{fig-cmds} the time evolution of the colour-magnitude
diagram (CMD) of the cluster is shown. Features such as the binary
main sequence, blue stragglers, and a collection of giants and white
dwarfs are clearly visible. The pre-main sequence stars begin above
the main sequence and descend down towards the main sequence. In the
more evolved colour-magnitude diagrams, a gap is noticeable in the
main sequence for the rw-pms and pz-pms simulations. A similar gap has
been observed in NGC 3603 \citep{EQZG98} at a mass of about 4
\msun. Since NGC 3603 is less than 5 Myr old, the gap should continue
to move down the zero age main sequence as the cluster ages. The gap
itself is a result of non-linearities in the mass-absolute magnitude
relation. These non-linearities arise from CNO burning that is
initially out of equilibrium in pre-main sequence stars
\citep{PB96}. Another notable feature of the CMDs is that the binary
main sequence is more sparsely populated in the pz-pms1 run. This is
another result of the increased number of early binary mergers, which
results in a reduced binary fraction in these runs.

\clearpage

\clearpage

\subsection{Local Stellar Properties}

The stellar populations in the various runs are quite different.
Pre-main sequence stars are quite numerous throughout the lifetime of
each of the clusters starting at the pre-main sequence. Aside from
that, there is a slight decrease in the number of stellar remnant and
giant stars. Tables \ref{stel_ms}, \ref{stel_pms} and \ref{stel_pms1}
show the time evolution of the various populations (single stars and
binaries) in the cluster. Here and throughout, `pms' stands for
pre-main sequence star, `ms' stands for main sequence star, `gs'
stands for giant star, and `rm' stands for stellar remnant. Each table
was created by averaging the results over three different realizations
of the initial conditions. The differences between the tables are due
to true differences between calculations, since each set of runs used
identical realizations of the initial snapshots, and only the initial
evolutionary state of the stars (zero age main sequence or pre-main
sequence) and the binary orbital properties were changed, as outlined
above.

Almost all of the features seen in tables \ref{stel_ms} --
\ref{stel_pms1} are attributable to either the difference in the
number of mergers or the shift in age that the entire population
experiences due to starting on the pre-main sequence. The fact that
the lower mass stars start their main sequence evolution later results
in a non-uniform shift in the population. It is interesting to note
that in many instances, the pz-ms runs sit in between the ps-pms
and rw-pms runs. For instance, the number of white dwarfs in the pz-ms
runs stays steadily between the rw-pms and the ps-pms runs. The reason
that there are so many white dwarfs in the pz-pms runs with respect to
the rw-pms runs is that the merger products from early in a pz-pms
cluster's lifetime will have a high mass, and thus tend to move more
quickly to the remnant stage. 

\clearpage

\begin{table}[t]
\caption{Population Evolution of pz-ms Runs}
\label{stel_ms}
\centering
\begin{tabular}{lrrrrrrrr}
\hline\hline
time [Myr]: & 0 & 100 & 200 & 400 & 600 & 800 & 1000 & 1200 \\ \hline
ms & 1024 & 1407.7 & 1374.3 & 1251.0 & 1109.0 & 931.7 & 763.3 & 610.7 \\
gs & 0 & 4.0 & 6.3 & 7.3 & 9.0 & 7.0 & 8.0 & 5.7 \\
rm & 0 & 7.0 & 14.0 & 27.0 & 39.7 & 48.0 & 51.0 & 53.3 \\
ms/ms & 1024 & 788.3 & 766.0 & 714.7 & 628.7 & 554.3 & 464.7 & 377.0 \\
ms/gs & 0 & 0.6 & 1.3 & 1.3 & 3.3 & 2.7 & 2.0 & 2.3 \\
ms/rm & 0 & 0.6 & 3.3 & 5.0 & 4.3 & 7.0 & 9.0 & 9.7 \\
gs/gs & 0 & 0.0 & 0.3 & 0.0 & 0.0 & 0.0 & 0.7 & 0.0 \\
gs/rm & 0 & 0.0 & 0.6 & 1.3 & 0.7 & 0.7 & 0.7 & 1.0 \\
rm/rm & 0 & 0.0 & 0.6 & 3.3 & 3.3 & 4.0 & 5.0 & 3.3 \\ \hline
\end{tabular}
\end{table}

\clearpage

\begin{table}[t]
\caption{Population Evolution of pz-pms Runs}
\label{stel_pms}
\centering
\begin{tabular}{lrrrrrrrr}
\hline\hline
time [Myr]: & 0 & 100 & 200 & 400 & 600 & 800 & 1000 & 1200 \\ \hline
pms & 1020.3 & 1395.3 & 1274.7 & 1071.0 & 868.7 & 684.7 & 524.3 & 379.7 \\
ms & 3.7 & 144.7 & 219.0 & 288.3 & 320.0 & 333.3 & 320.3 & 302.3 \\
gs & 0.0 & 3.7 & 5.7 & 6.3 & 7.7 & 8.3 & 8.7 & 7.0 \\
rm & 0.0 & 7.0 & 15.3 & 31.3 & 43.3 & 53.0 & 58.3 & 60.7 \\
pms/pms & 1020.6 & 600.7 & 553.3 & 473.3 & 392.7 & 322.3 & 258.3 & 200.3 \\
pms/ms & 2.7 & 34.7 & 56.0 & 74.3 & 86.0 & 85.3 & 85.0 & 75.7 \\
pms/gs & 0.0 & 0.3 & 0.0 & 0.3 & 0.0 & 1.0 & 1.0 & 0.3 \\
pms/rm & 0.0 & 0.0 & 0.3 & 0.3 & 0.3 & 0.3 & 1.0 & 1.7 \\
ms/ms & 0.7 & 23.7 & 28.7 & 45.0 & 55.0 & 59.3 & 61.7 & 62.0 \\
ms/gs & 0.0 & 0.3 & 1.7 & 0.7 & 2.7 & 1.0 & 0.7 & 1.3 \\
ms/rm & 0.0 & 0.0 & 1.7 & 3.0 & 4.0 & 5.3 & 5.7 & 5.7 \\
gs/gs & 0.0 & 0.3 & 0.0 & 0.0 & 0.3 & 0.0 & 0.0 & 0.0 \\
gs/rm & 0.0 & 0.0 & 0.7 & 0.7 & 0.3 & 0.7 & 0.3 & 0.3 \\
rm/rm & 0.0 & 0.0 & 1.7 & 2.0 & 1.7 & 2.0 & 1.7 & 1.3 \\ \hline
\end{tabular}
\end{table}

\clearpage

\begin{table}[t]
\caption{Population Evolution of rw-pms Runs}
\label{stel_pms1}
\centering
\begin{tabular}{lrrrrrrrr}
\hline\hline
time [Myr]: & 0 & 100 & 200 & 400 & 600 & 800 & 1000 & 1200 \\ \hline
pms & 1020.3 & 1450.0 & 1346.7 & 1148.3 & 936.3 & 735.7 & 566.0 & 471.7 \\
ms & 3.7 & 127.3 & 192.3 & 255.3 & 296.3 & 301.3 & 284.3 & 291.3 \\
gs & 0.0 & 4.3 & 5.0 & 3.7 & 7.7 & 7.3 & 8.3 & 8.7 \\
rm & 0.0 & 7.0 & 15.3 & 28.7 & 35.7 & 43.3 & 48.3 & 51.0 \\
pms/pms & 1020.6 & 643.0 & 600.0 & 512.3 & 418.7 & 334.7 & 262.3 & 226.7 \\
pms/ms & 2.7 & 44.3 & 61.0 & 87.3 & 95.0 & 104.3 & 100.3 & 97.3 \\
pms/gs & 0.0 & 0.0 & 0.3 & 1.0 & 0.3 & 1.7 & 0.7 & 0.3 \\
pms/rm & 0.0 & 0.0 & 0.3 & 0.7 & 2.0 & 1.7 & 2.7 & 3.3 \\
ms/ms & 0.7 & 22.3 & 32.7 & 47.0 & 56.3 & 64.3 & 67.7 & 66.7 \\
ms/gs & 0.0 & 0.0 & 2.3 & 1.7 & 2.3 & 0.3 & 1.0 & 1.0 \\
ms/rm & 0.0 & 1.3 & 1.3 & 4.7 & 6.0 & 7.7 & 6.7 & 6.3 \\
gs/gs & 0.0 & 0.0 & 0.0 & 0.0 & 0.3 & 0.0 & 0.0 & 0.0 \\
gs/rm & 0.0 & 0.0 & 0.0 & 0.3 & 1.3 & 0.3 & 0.3 & 1.0 \\
rm/rm & 0.0 & 0.3 & 1.0 & 2.3 & 3.0 & 4.7 & 5.0 & 4.0 \\ \hline
\end{tabular}
\end{table}

\clearpage

\begin{figure*}
\plottwo{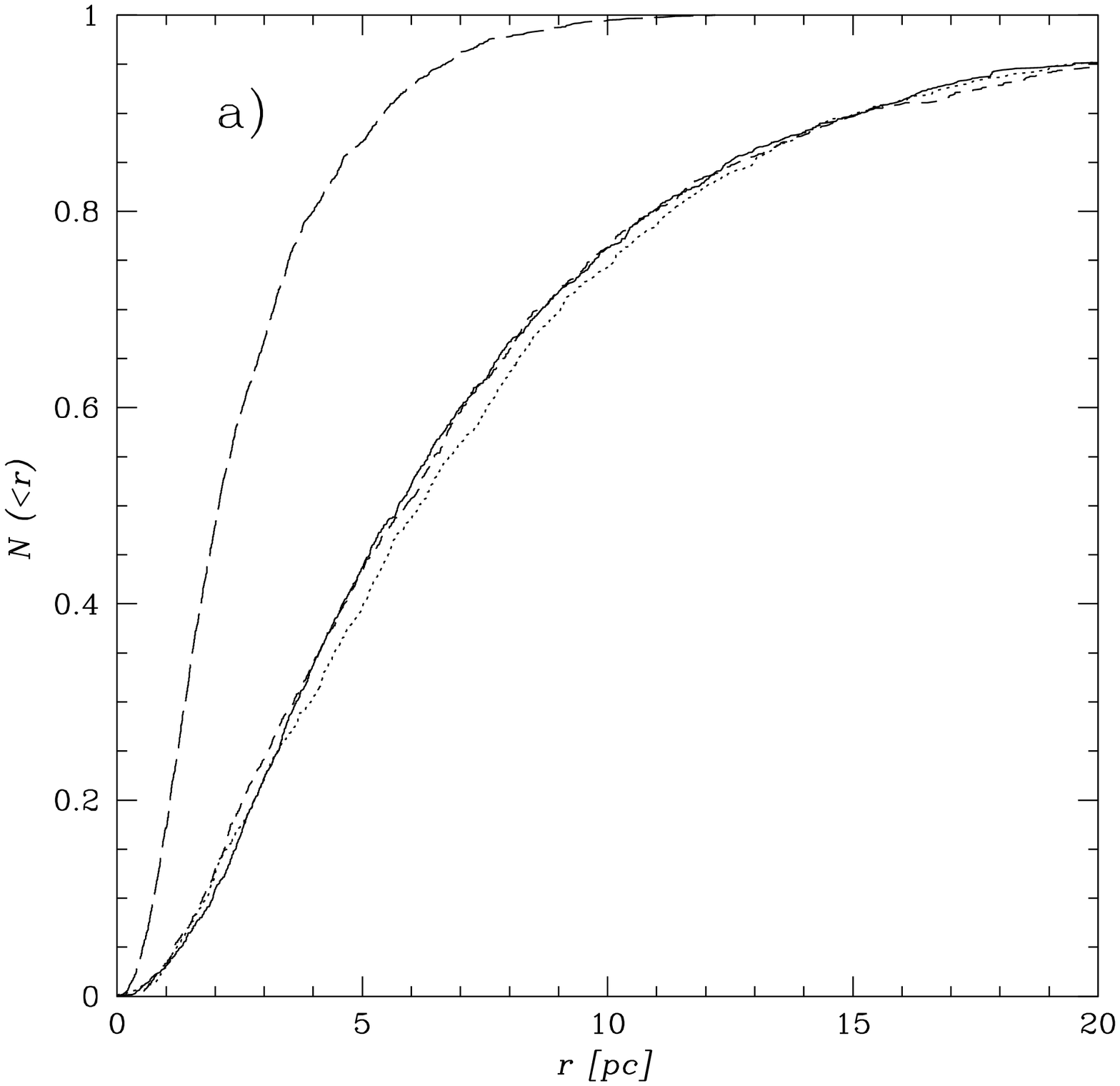}{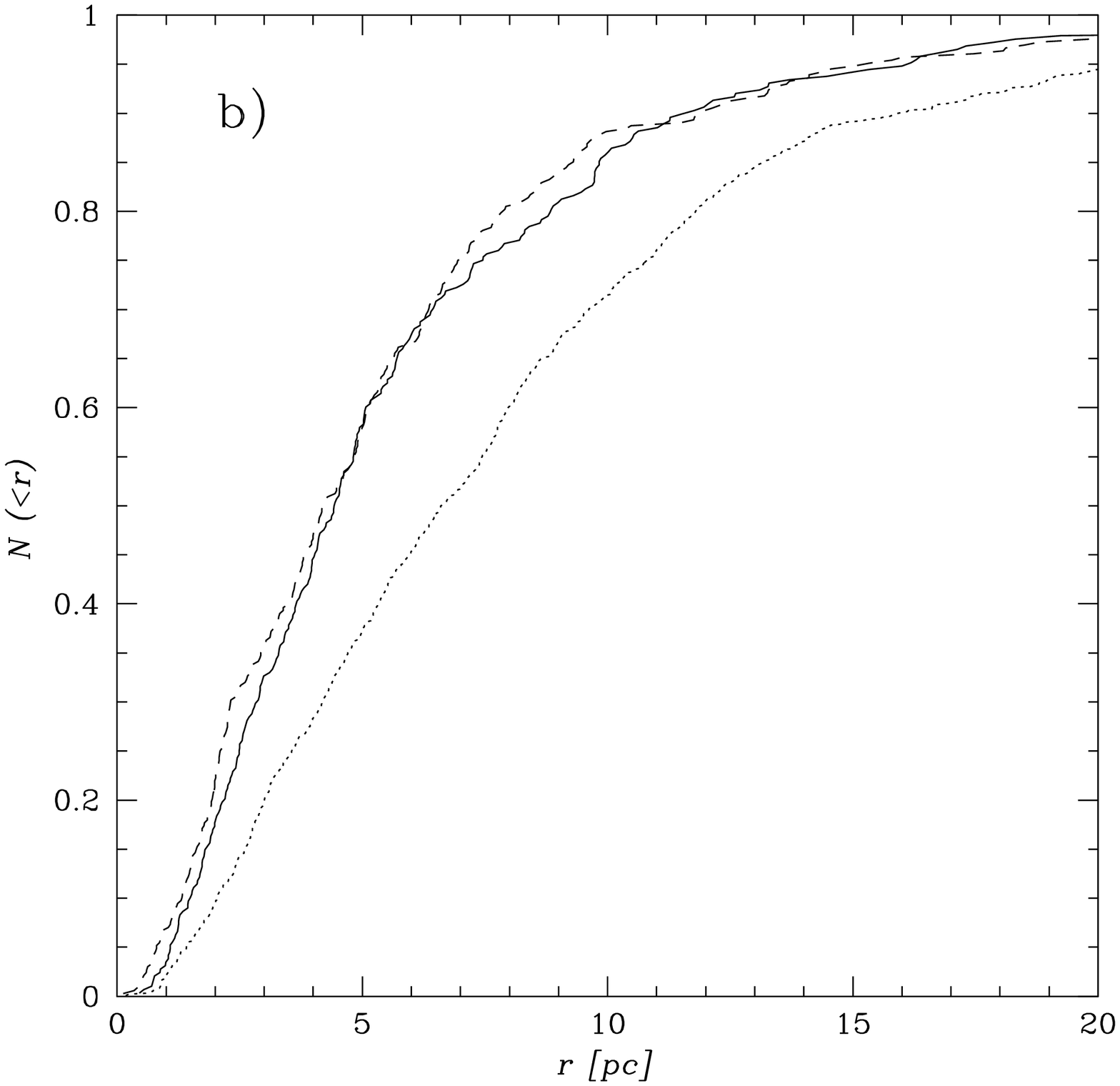}
\caption{Radial distributions for runs pz-ms1 (dotted line) , pz-pms1 
(short-dashed line), and rw-pms1 (solid line) at 600 Myr. Panel a)
gives the radial distribution for all objects, while panel b) shows
the radial distribution for main sequence stars only. The long-dashed
line in panel a) shows the initial radial distribution of all
objects. \label{popdist}}
\end{figure*}

In general, the different stellar populations end up being similarly
distributed throughout the cluster. Figure \ref{popdist} shows the
cumulative radial distribution of all stars (panel a) and for main
sequence stars only (panel b) at 600 Myr for the first run of each of
the series. The radial distribution of all stars is almost identical
for all three simulations, and this is also true for the individual
populations of binary stars, giants, remnants, and pre-main sequence
stars where relevant. The biggest difference is in the main sequence
stars, as shown in panel b) of figure \ref{popdist}. The main sequence
stars are significantly more centrally concentrated in the two runs
with pre-main sequence evolution than in the pz-ms run. The main
reason for this is that the lowest mass stars are still on the
pre-main sequence. Therefore, the mean mass of main sequence stars is
higher than that in the pz-ms run, and the population is more
centrally concentrated.

\clearpage

\clearpage

\begin{figure*}
\epsscale{0.4}
\plotone{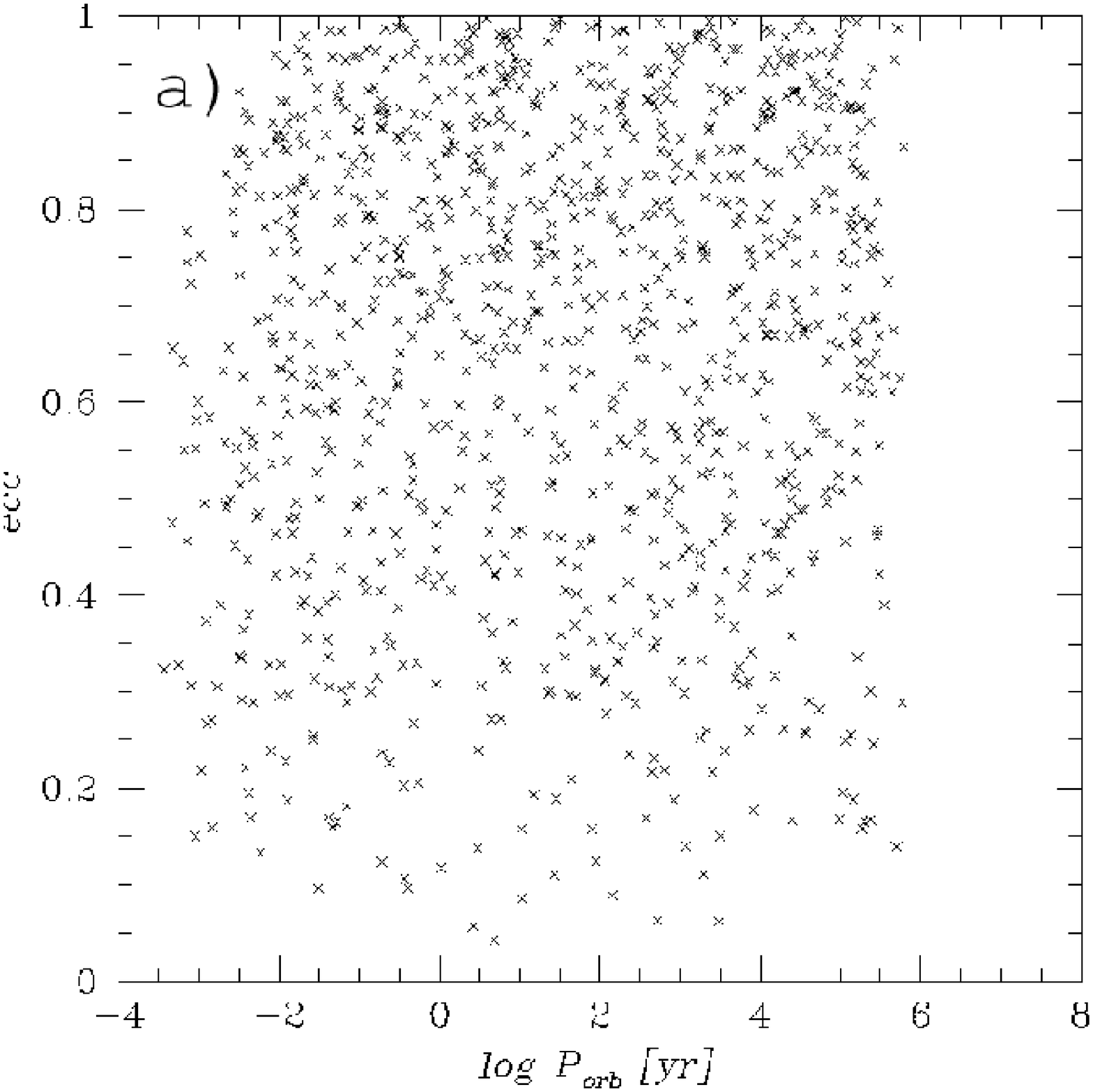}
\plotone{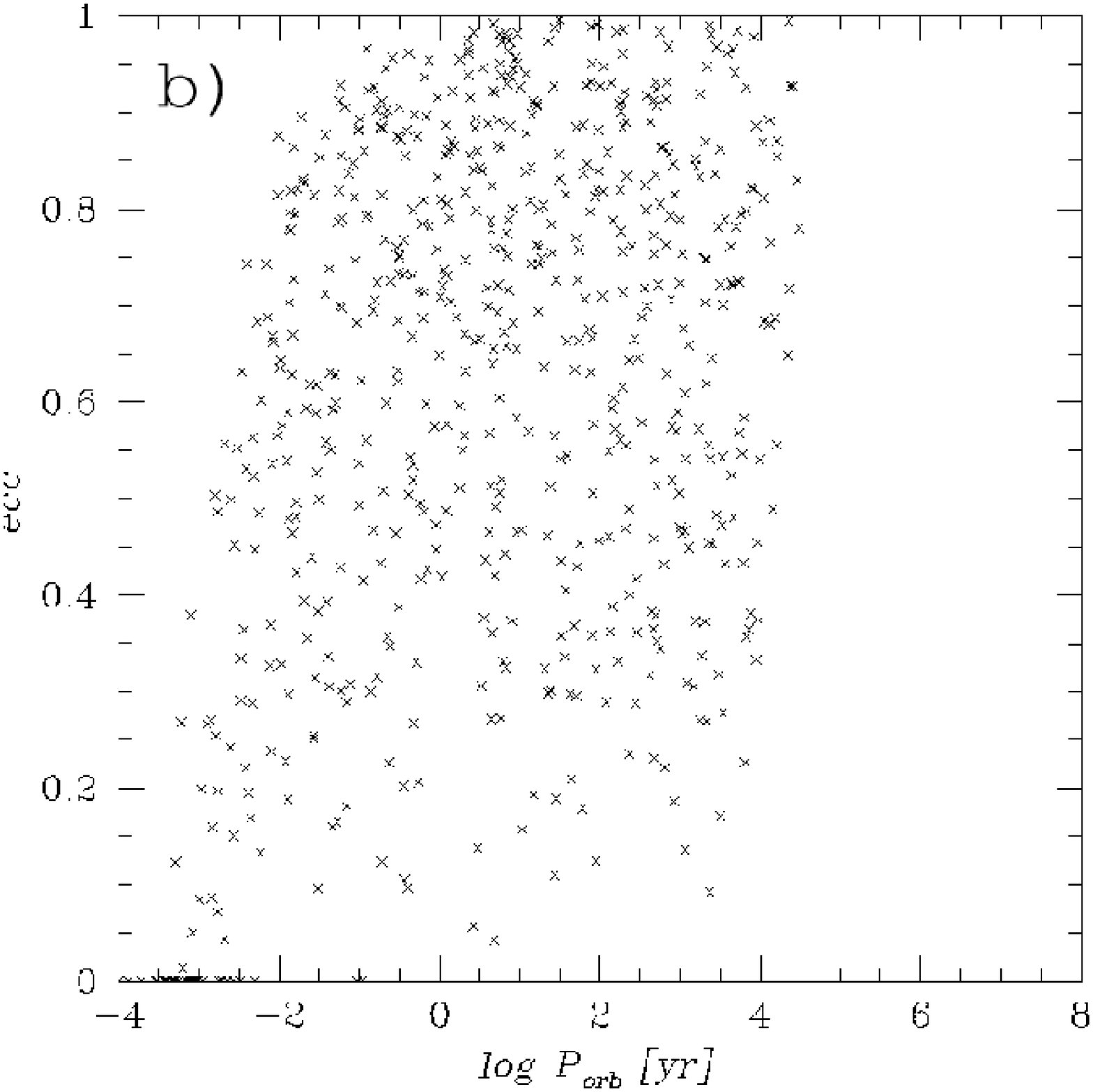}
\plotone{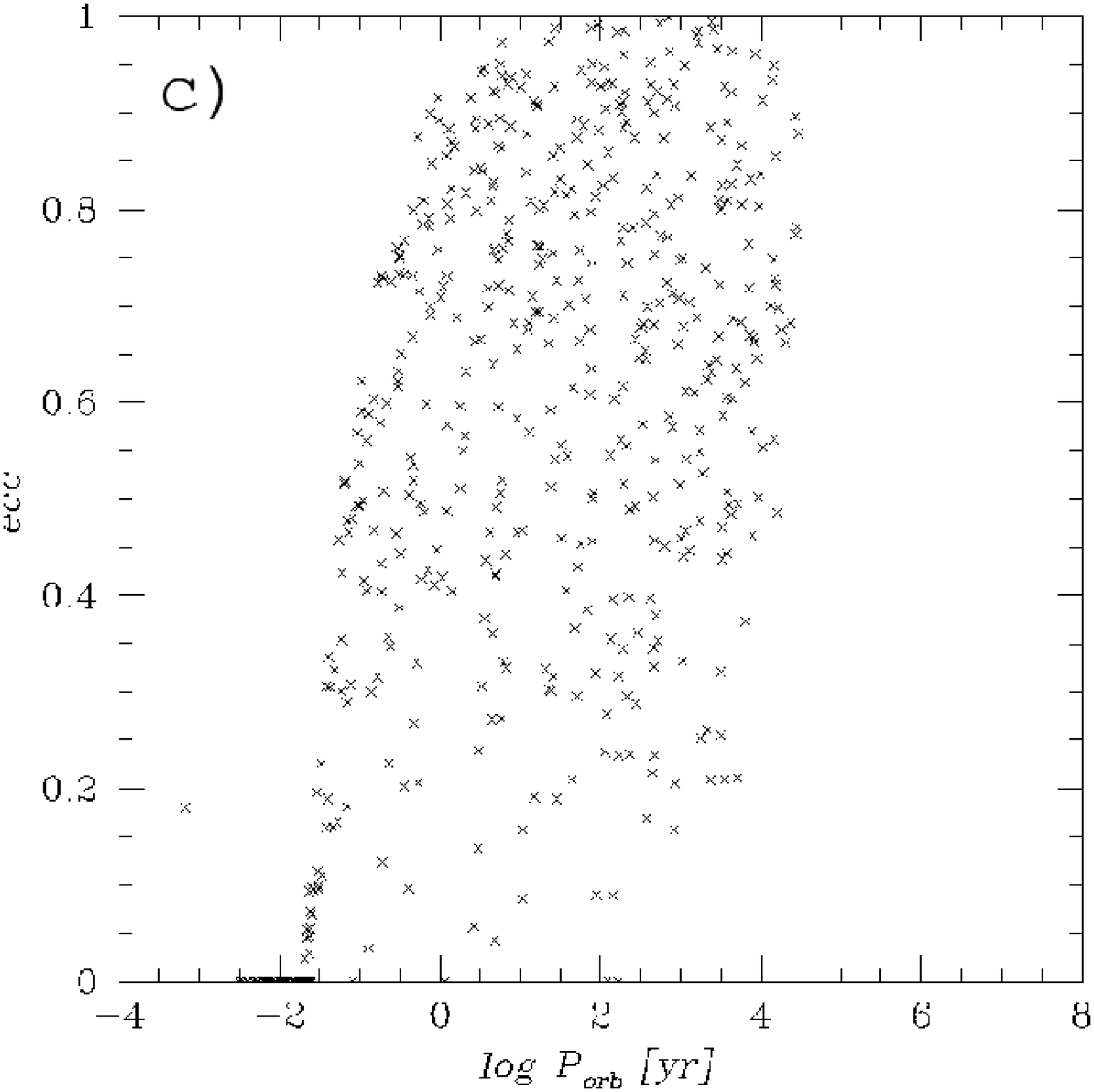}
\epsscale{1}
\caption{Eccentricity vs. orbital period of binaries for pz-ms1 and
pz-pms1. Shown are the initial parameters shared by both runs (panel a),
and the parameters at 600 Myr for pz-ms1 (panel b) and pz-pms1
(panel c).\label{binorbspz}
 }
\end{figure*}

\begin{figure*}
\plottwo{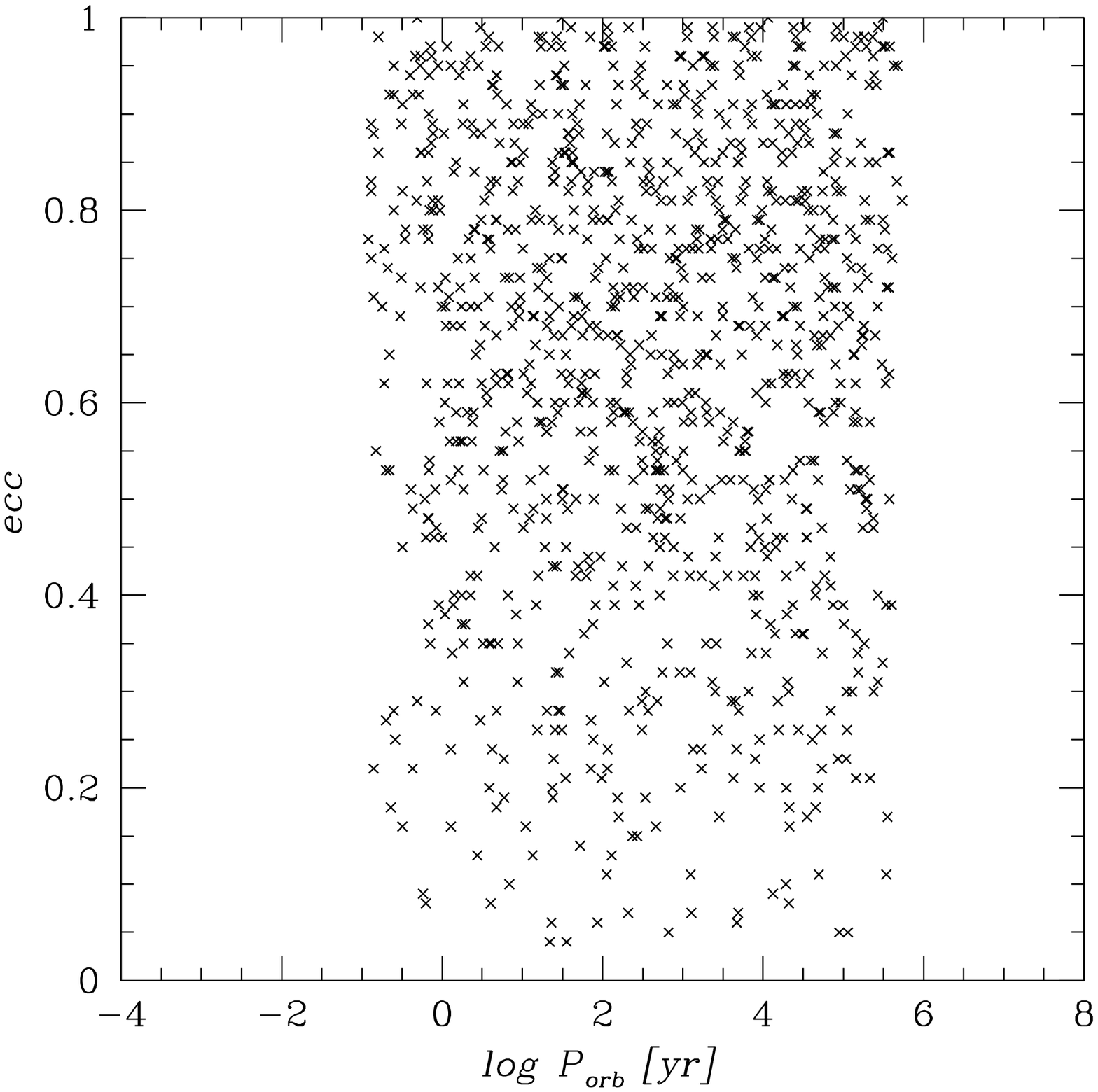}{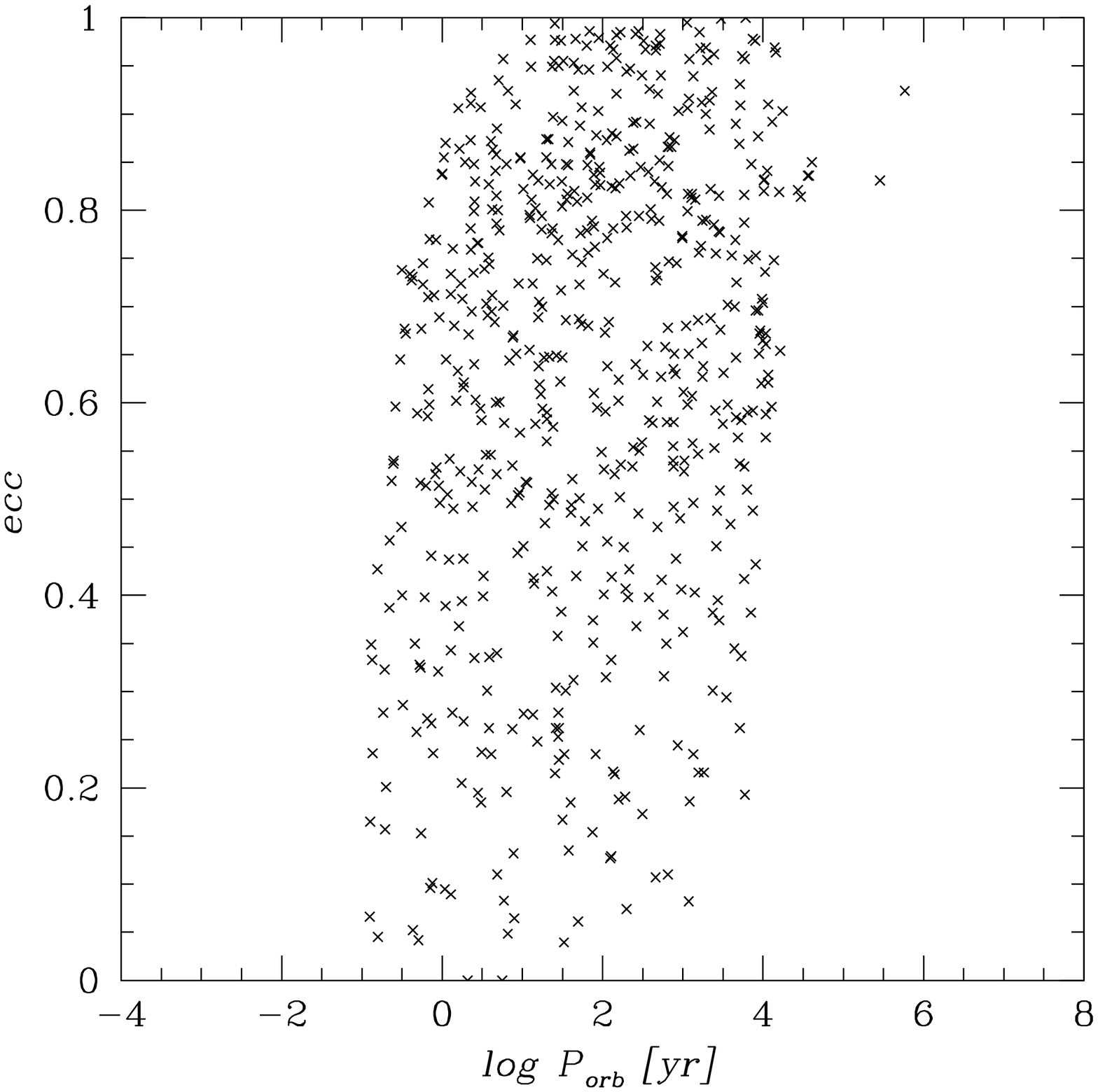}
\caption{Eccentricity versus orbital period of binaries for rw-pms1.
Shown are the parameters at 0 Myr (left) and 600 Myr
(right).\label{binorbsrw}}
\end{figure*}

Figures \ref{binorbspz} and \ref{binorbsrw} show the binary orbital
parameters (semi-major axis $a$ and eccentricity $e$) of all stars at
the start of the simulation and at 600 Myr for runs using the original
binary orbits and using the more realistic binary orbits
respectively. For the pz-pms simulation it is evident that almost all
of the binaries with a small orbital period have merged or
circularized by 600 Myr. For the rw-pms model almost no
circularization occurs. This is again attributable to the fact that
the pre-main sequence stars are contracting and are therefore rarely
in a period where tidal effects are important. As is expected, all of
the softer binaries (those with orbital periods larger than $10^4$ yr)
have been broken up in all models by 600 Myr. In addition, the
pre-main sequence systems that are not initially in contact will only
experience Roche lobe overflow after the terminal age main sequence
when the star ascends the giant branch. Only then will these stars be
larger than they were in their initial stage. Stars that are
considered to begin their lives on the zero-age main sequence expand
as they evolve, so there are more systems which experience Roche lobe
overflow on the main sequence in the pz-ms runs.

We can directly compare the effects of including the pre-main sequence
phase on binary evolution by making use of the information shown in
figure \ref{binorbspz}. Both the pz-ms and the pz-pms runs started
with the same 1024 binaries (those shown in the first panel). After
600 Myr, the pz-ms run has 639 binaries while the pz-pms run has only
545. However, only 426 systems are still in common between the two
simulations. Of those 426, 300 have evolved in exactly the same manner
and have exactly the same orbital elements. Many of those systems
consisted of a pair of massive stars, and so were unaffected by the
inclusion of the pre-main sequence phase in our simulations. The
evolution of the other 126 systems is quite enlightening. We can
calculate the total change in semi-major axes ($\Delta a = \Sigma_{\rm
all binaries} a_{\rm pz-ms} - a_{\rm pz-pms}$) and the equivalent
quantity for eccentricity for the systems that are in common between
the two simulations. The total change in semi-major axis is relatively
small ($\Delta a = -150 $ A.U. summed over 126 systems) and the
differences are almost equally probable to be positive or
negative. For comparison, the maximum binary semi-major axis in our
simulations was $10^6$ \rsun, or $\sim 4600$ A.U. The situation for
eccentricity is quite different: $\Delta e = 45$. This very large
number shows that the binaries in the ps-pms simulation are much more
likely to have substantially lower eccentricities than the pz-ms
run. This can be attributed to a phase of tidal circularization that
occurs while the stars are on the pre-main sequence, and can be seen
in the lack of high eccentricity systems in the third panel of figure
\ref{binorbspz}.

It has been known for some time that there is a correlation between
the age of a stellar population and the binary period below which all
binaries in that population are circularized. This correlation is
based on the efficiency of tidal circularization. However, the current
models for circularization do not agree with all the available data
(see \citet{MM05} for a recent review and investigation of this
problem). We have shown in figure \ref{binorbspz} that the tidal
circularization period can differ by more than two orders of magnitude
for two different assumptions about the properties of the initial
starting point of the stars. Observations suggest that the tidal
circularization period in the Hyades is $3.2 \pm 1.2$ days, or $\log
P_{\rm orb}({\rm yrs}) \sim -2$ \citep{MM05}. The pz-pms run is a much
better fit to the data than the pz-ms run. In the rw-pms run, however,
there are no binaries (either initially or after 600 Myr) with orbital
periods that low. Once again, it is clear that understanding the
initial binary properties on the pre-main sequence is crucial.

\clearpage

\clearpage

\clearpage

\begin{figure*}
\plotone{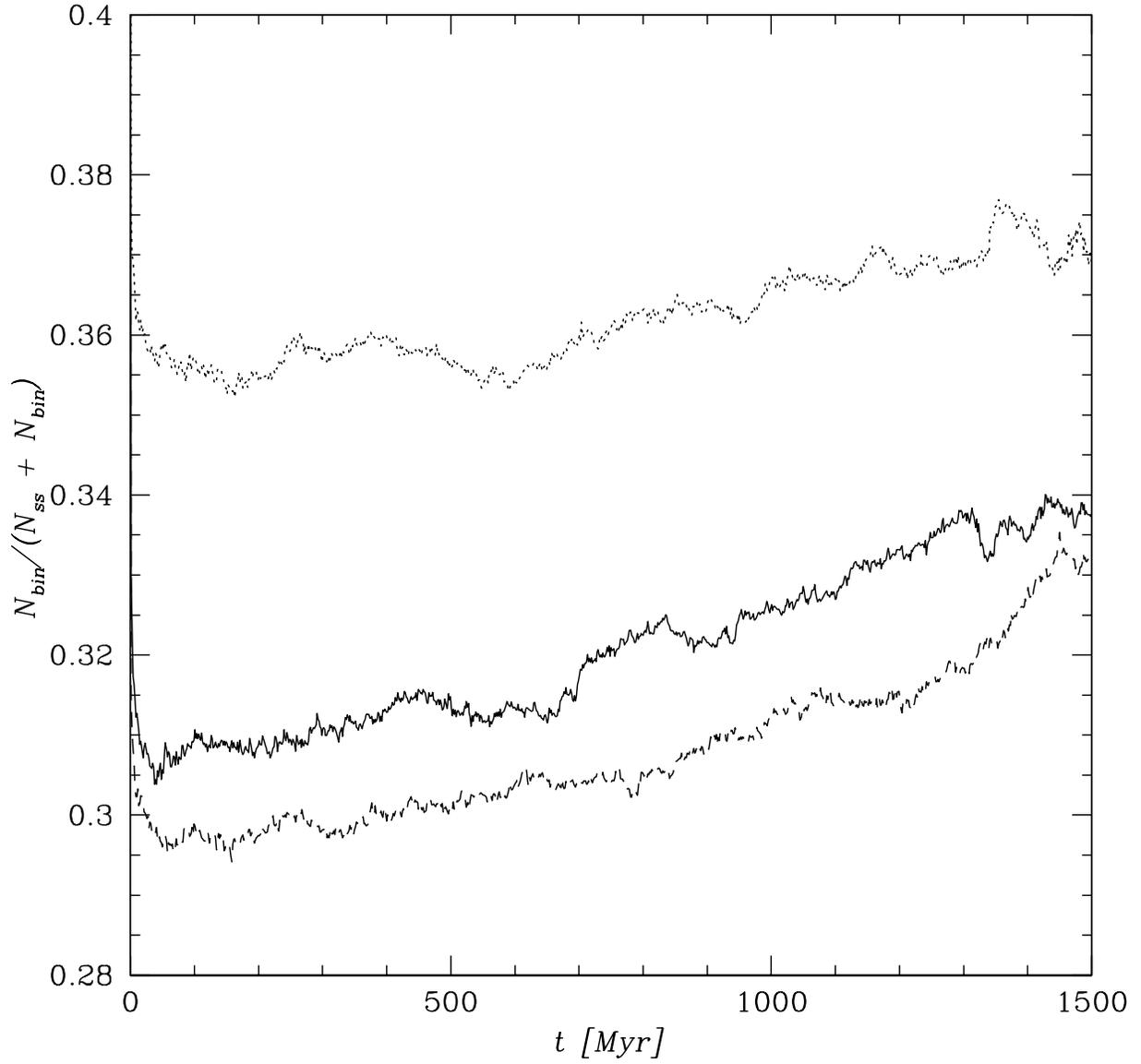}
\caption{Binary fraction versus time for all binaries. Shown are pz-ms
(dotted), rw-pms (solid),and pz-pms (dashed), averaged over all
runs.\label{binfrac}}
\end{figure*}

Throughout the pz-pms and rw-pms runs, the binary fraction stays below
that of the pz-ms runs, as seen in figure \ref{binfrac}. The initial
dip for the pz-pms runs is a result of the large number of mergers
that occur within the first Myr after the simulation begins. For
the rw-pms runs, the binary orbits were biased towards a higher
fraction of soft binaries. This leads to many multiple systems
breaking up into single stars early in the cluster's life. The binary
fraction in both pre-main sequence simulations increases slightly with
time because binaries are formed through stellar interactions in the
core of the cluster, and because single stars are preferentially
ejected from the cluster. The binary fraction within the inner 1 pc of
the rw-pms cluster reaches about 75\% after 600 Myr, and falls off
steeply in the outskirts of the cluster. The number of very hard
binaries ($E < -1000$\,kT) is very small in the rw-pms run, making up no
more than 6\% of the binary fraction, compared to about twice that for
the pz-ms run. Detailed observations of the binary properties
in open clusters may therefore be able to constrain the
initial binary properties in the cluster, if the observations are
sufficiently complete and sufficiently precise to determine binary
orbital parameters.

\clearpage

\clearpage

The most obvious and most important difference between runs of the
pz-pms type and those of the other types are the number of mergers. In
table \ref{mergers} we give the total numbers of mergers seen in all
simulations, sorted by kind of simulation. Because many of the
binaries were in contact for the initial pz-pms conditions, the first
timestep contained a very large number of mergers. On the other hand,
the number of mergers for the rw-pms simulations was lower than that
of the pz-ms simulations. This is because the binaries which included
pre-main sequence stars started with large separations. The pre-main
sequence stars then contracted towards their zero age main sequence
radii, resulting in an increasing ratio of stellar radius to Roche
lobe radius. These binaries will take longer to evolve towards Roche
lobe contact than the binaries in the pz-ms runs, and Roche lobe
overflow will happen after the star has started to ascend the giant
branch. This is another indication that the most important thing to
consider when including the pre-main sequence phase of stellar
evolution is the initial parameters of binary systems in the cluster.

\clearpage

\begin{table}
\caption{Total number of different kinds of mergers for each run type.}
\label{mergers}
\begin{tabular}{lrrr}
\hline\hline
Merger Type & PZ-MS & PZ-PMS & RW-PMS \\
\hline
pms/pms & 0  & 441 & 0 \\
pms/ms  & 0  & 0   & 1 \\
pms/gs  & 0  & 5   & 7 \\
ms/ms   & 86 & 1   & 1 \\
ms/gs   & 8  & 1   & 0 \\
ms/rm   & 4  & 2   & 1 \\
gs/gs   & 1  & 0   & 0 \\
gs/rm   & 7  & 7   & 1 \\
rm/rm   & 2  & 2   & 1 \\ 
\hline
\end{tabular}
\end{table}

\clearpage
\section{Summary and Discussion}

In this paper, we investigated the effects of allowing stars to begin
their lives on the pre-main sequence in dynamical simulations. We
added pre-main sequence evolutionary tracks which begin at the
deuterium-burning birthline, and end at the zero-age main
sequence. Stars with masses less than 7 \msun\ in the dynamical
simulations have their radii, luminosities, temperatures, and other
properties determined by these tracks starting from the $T=0$ point of
the dynamical simulation. We compared these simulations with
standard ones in which all stars begin their lives on the zero age
main sequence at $T=0$.

The dominant characteristic which distinguishes between our different
initial conditions can be traced back to the number of mergers in the
early stage of the cluster evolution. In the pz-pms runs, a lot of the
binaries begin in Roche lobe contact, since the binary orbits are
based on zero age main sequence radii, and pre-main sequence stars
have much larger radii initially. These early pre-main
sequence/pre-main sequence mergers show up as a decrease in the total
number of stars, but an increase in average mass. This effect changes
the mass function of the cluster, not only initially, but the effect
continues to be noticed for quite a large portion of the cluster's
lifetime (many tens of initial half-mass relaxation times). On the
other hand, since the stars in the rw-pms runs start at a greater
separation, and since the initial pre-main sequence radius is the
largest radius for a given star until it becomes a giant star, there
is an absence of mergers in these runs. This increases the binary
fraction, and affects the mass function as well.

In spite of the initial drop of total number of objects in the pz-pms
models, the time evolution of the total mass is not noticeably affected
by the starting point of the stars. Similarly, the total number of
stars in both pms simulations seem to decrease at a slower rate than
their pz-ms counterparts. This is due to a smaller number of
escapers ejected from the cluster through an encounter. The
inclusion of pre-main sequence evolution causes the stellar
interactions to be less violent. During a binary star/single star
interaction, accretion during the stellar encounter affects the
ensuing binary parameters which in turn affects the resulting
ejection velocity. Indeed, the merger of close binaries (as in the
pz-pms runs) or the absence of close binaries in the rw-pms runs
should affect the cluster in the same way since binaries act as a heat
sink for the total energy of the cluster. Since the predominant form
of binary that stars in these models will encounter is not very hard,
the heat sinks of the cluster can absorb more, thus decreasing the
energy available for other purposes.

In the rw-pms models very little circularization is observed. This is
because the stars are contracting away from each other, and therefore
cannot maintain close proximity as required for circularization. The
pz-pms runs contain the largest number of circularized binaries and
has a tidal circularization period which best matches observations of
the Hyades.

Another item of note is the frequency of the different types of
mergers. Almost all of the mergers were due to the normal binary
evolution in which a star's radius becomes large than its Roche lobe.
The two most frequent mergers in the pms runs are pre-main
sequence/pre-main sequence and pre-main sequence/giant star mergers
(with the latter being more dominant in the rw-pms case).
Hydrodynamic simulations of such collisions \citep{LS05} suggest that
the result will be a larger pre-main sequence star in the former case,
and a larger giant star in the latter case.

There is a significant amount of future work that could be done in
this area and could involve simulating other cluster configurations
(for instance, with more stars or in different tidal field). Since
pre-main sequence stars in very young clusters have been studied
(e.g., \cite{EQZG98}), a focus on the first 100 Myr of a simulation
could be insightful. Another avenue of research would be to perform
population synthesis in order to better determine which initial
distribution of binary parameters will result in the observed
distribution of these parameters, along the lines of \citet{K95}. Both
the choices for initial binary orbits in the rw-pms and pz-pms runs
were quite simplistic and should be improved upon.

There are two noteworthy pieces of astrophysics that we have
completely neglected from these simulations. Both are particularly
relevant to the study of young clusters, particularly clusters
significantly younger than the 600 Myr Hyades analogue we concentrated
on here. The first topic is the inclusion of the gas out of which the
stars in these clusters formed. We know from observations of star
forming regions that there is a reasonably long period of time when
both stars and gas co-exist, and are presumably interacting
dynamically. We neglect that stage completely by starting with a
stars-only King model. The second topic is the effect of circumstellar
(and circumbinary) disks around the young stars in the
cluster. Interactions between stars with disks should be slightly
different than interactions between stars without disks, and could
modify our results. In this paper, we have demonstrated that the
careful treatment of cluster initial conditions is important, and
there are clearly other avenues for improvement as well.

Our conclusion from this study is that inclusion of the pre-main
sequence phase of stellar evolution is critical for any simulations
that wish to understand the properties of binary stars in stellar
systems. The choice of initial binary parameters and their subsequent
evolution are strongly modified by the properties of pre-main sequence
stars. 

\acknowledgments

RW acknowledges support from SHARCNet. AS is supported by NSERC and
the Canadian Foundation for Innovation. SPZ acknowledges support from
the Dutch Royal Academy of Science (KNAW), Dutch Organization for
Scientific Research (NWO) and Dutch Research School for Astrophysics
(NOVA).

\end{document}